\documentclass[acmsmall]{acmart}

\usepackage[english]{babel}
\usepackage{blindtext,xcolor}
\usepackage{wrapfig}
\usepackage[utf8]{inputenc}
\usepackage{microtype}
\usepackage{wasysym}

\usepackage{enumitem}
\usepackage{graphicx}
\usepackage{array}
\usepackage{balance}
\usepackage{siunitx}
\usepackage{booktabs}
\usepackage{xparse}
\usepackage{tabularx}
\usepackage{url}
\usepackage[font=small]{caption}
\usepackage{subcaption}
\usepackage{makecell}
\usepackage{xspace}
\usepackage{multirow}
\usepackage{mathtools}
\usepackage{fancyvrb}
\usepackage{wrapfig}
\usepackage[ruled,vlined]{algorithm2e}

\newcommand{\lhh}{HitchHiking\xspace}
\newcommand{\SD}{San Diego\xspace}
\newcommand{\Stanford}{Stanford\xspace}

\usepackage{balance}
\usepackage{url}

\usepackage{breakurl}
\renewcommand\footnotetextcopyrightpermission[1]{}
\acmYear{2023}\copyrightyear{2023}
\acmConference[]{Pre-Print}{}
\begin{document}

\title{Democratizing LEO Satellite Network Measurement}
%\subtitle{\color{red}[PRE-PRINT]}
%\author{Paper \# 93}

\makeatletter
\let\@authorsaddresses\@empty
\makeatother

\author[]{Liz Izhikevich}
\affiliation[]{Stanford University}
\email{lizhikev@stanford.edu}

\author[]{Manda Tran}
\affiliation[]{Stanford University}

\author[]{Katherine Izhikevich}
\affiliation[]{UC San Diego}

\author[]{Gautam Akiwate}
\affiliation[]{Stanford University}

\author[]{Zakir Durumeric}
\affiliation[]{Stanford University}
%\email{zakir@cs.stanford.edu}

%make the title area
\renewcommand{\shortauthors}{L. Izhikevich et~al.}

\begin{abstract}

Low Earth Orbit (LEO) satellite networks are quickly gaining traction with promises of impressively low latency, high bandwidth, and global reach.
However, the research community knows relatively little about their operation and performance in practice. 
The obscurity is largely due to the high barrier of entry for measuring LEO networks, which requires deploying specialized hardware or recruiting large numbers of satellite Internet customers. 
In this paper, we introduce \lhh, a methodology that democratizes global visibility into LEO satellite networks.
\lhh builds on the observation that Internet-exposed services that use LEO Internet can reveal satellite network architecture and performance, bypassing the need for specialized hardware.
We evaluate \lhh against ground truth measurements and prior methods, showing that it provides more coverage and accuracy.
With \lhh, we complete the largest study to date of Starlink network latency, measuring over 2,400~users across 13~countries. 
We uncover unexpected patterns in latency that surface how LEO routing is more complex than previously understood.
Finally, we conclude with recommendations for future research on LEO networks.

\end{abstract}

\maketitle
\thispagestyle{empty}
\section{Introduction}

% The increasing adoption of satellite-based Internet Service Providers (ISPs) has also seen increasing numbers of satellite terminals and base stations exposed, not always voluntarily, on the public Internet. 

Low Earth Orbit (LEO) satellite networks
%are a new architecture for Internet connectivity that 
promise to decrease latency and increase Internet coverage
beyond the means of terrestrial Internet.
Used by at least 2~million people as of September 2023~\cite{spaceX-2mil}, LEO networks have proven immensely useful in natural disaster areas~\cite{natural_disaster}, 
underprivileged schools~\cite{under_prvlgd_schools}, war zones~\cite{ukrainewar}, and remote research labs~\cite{remote_labs}.
To improve Internet access for those that rely on LEO networks, researchers have begun to propose improvements to their latency~\cite{bhattacharjee2023laser}, routing~\cite{bhosale2023characterization,wang2023reliability}, and security~\cite{giuliari2021icarus}.
%nderstand LEO network architecture~\cite{hypatia} and performance~\cite{imcdish22,ma2022network,kassem2022browser}

Unfortunately, researchers today face a barrier when understanding how LEO networks operate in practice: to collect real data, one must acquire expensive specialized satellite hardware or recruit volunteers that use the satellite network. %hardware-equipped. %, in order to communicate with satellites.
Consequently, the community's study of LEO networks has been limited to a small number of vantage points~\cite{kassem2022browser,ripe_atlas,imcdish22,ma2022network} and unvalidated theoretical models~\cite{hypatia,giuliari2021icarus,bhosale2023characterization,wang2023reliability,bhattacharjee2023laser}.
If researchers plan to help design and protect LEO networks, it behooves us to lower the barrier to acquiring data, 
such that we have an accurate understanding of how the worldwide LEO network ecosystem works in practice.

In this paper, we introduce \lhh, a methodology to actively measure LEO satellite network characteristics at scale. \lhh builds on the key insight that probing publicly exposed devices on LEO satellite networks can reveal characteristics of the underlying satellite network. Crucially, the \lhh methodology requires no specialized hardware or painstaking recruitment, lowering the barrier to collect real data from LEO networks worldwide. \lhh works in three steps (1) find measurable endpoints that use LEO satellites for Internet connectivity, (2) identify where in the network path LEO satellites are used, and 
(3) measure (i.e., ``hitchhike'') the satellite link. 

We use \lhh to conduct the largest LEO network measurement to date---measuring the latency of 2.4K LEO satellite customers across 13~countries---of Starlink~\cite{starlink}, the largest LEO network.
%links that belong to the only commercial LEO network that sells to individual consumers: SpaceX-Starlink
%\lhh measures over 1.5k LEO links across 13~countries, an order of magnitude more than the leading alternative~\cite{ripe_atlas}. 
We show that \lhh collects nearly identical latency data compared to the ground truth observed by a physical Starlink dish. 
%\lhh is the first methodology that does not require specialized hardware to gain a worldwide data-driven perspective about LEO networks. 
We then use \lhh's measurement---and
%of users across
%\lhh's ability to measure users across farms, boats, and remote locations, 
validation by Starlink---to surface the three primary ways in which Starlink's network does not operate as prior work assumed.
First, sustained peaks of latency are not due to changes in satellite location.
Second, customer latency is bounded by the availability of a nearby Point of Presence (POP). 
Third, the use of inter-satellite links significantly increases routing path lengths and latency.

% illuminates that .
% .
% Furthermore, we discover while inter-satellite links---a LEO routing technology with latency below terrestrial fiber---likely do minimize user to ground station routing, they significantly increase the routing path between the ground station to POP. 

Our work illustrates that a diverse set of perspectives are needed to understand the 
unique routing and latency properties of LEO networks. For researchers interested in studying LEO networks, the \lhh methodology we introduce provides accessibility to those desiring experiment flexibility and coverage when collecting real data. To lower the barrier for future LEO research, we are releasing the \lhh pipeline for measuring Starlink latency and the data \lhh collects under the Apache 2.0 license.

\section{Background}
\label{sec:background}

Satellite Internet has existed for over 20~years~\cite{anik_f2}. 
However, early Internet-providing satellites were large, expensive, and geostationary (i.e., fixed with respect to a position on the Earth). Geosynchronous equatorial orbit (GEO) satellites orbit over 22,000~miles from earth. While this long distance brings wide coverage, it comes at the expense of latency. Even today, the minimum round trip time (RTT) for a packet to route through a GEO satellite is 480~ms, physically bounded by the speed of light. 
In response to lower latency requirements---and cheaper satellite technology---a new class of satellite networks emerged: Low Earth Orbit (LEO) satellite networks. 

\begin{wrapfigure}{l}{0.5\textwidth}
  %\centering
    \includegraphics[scale=0.5]{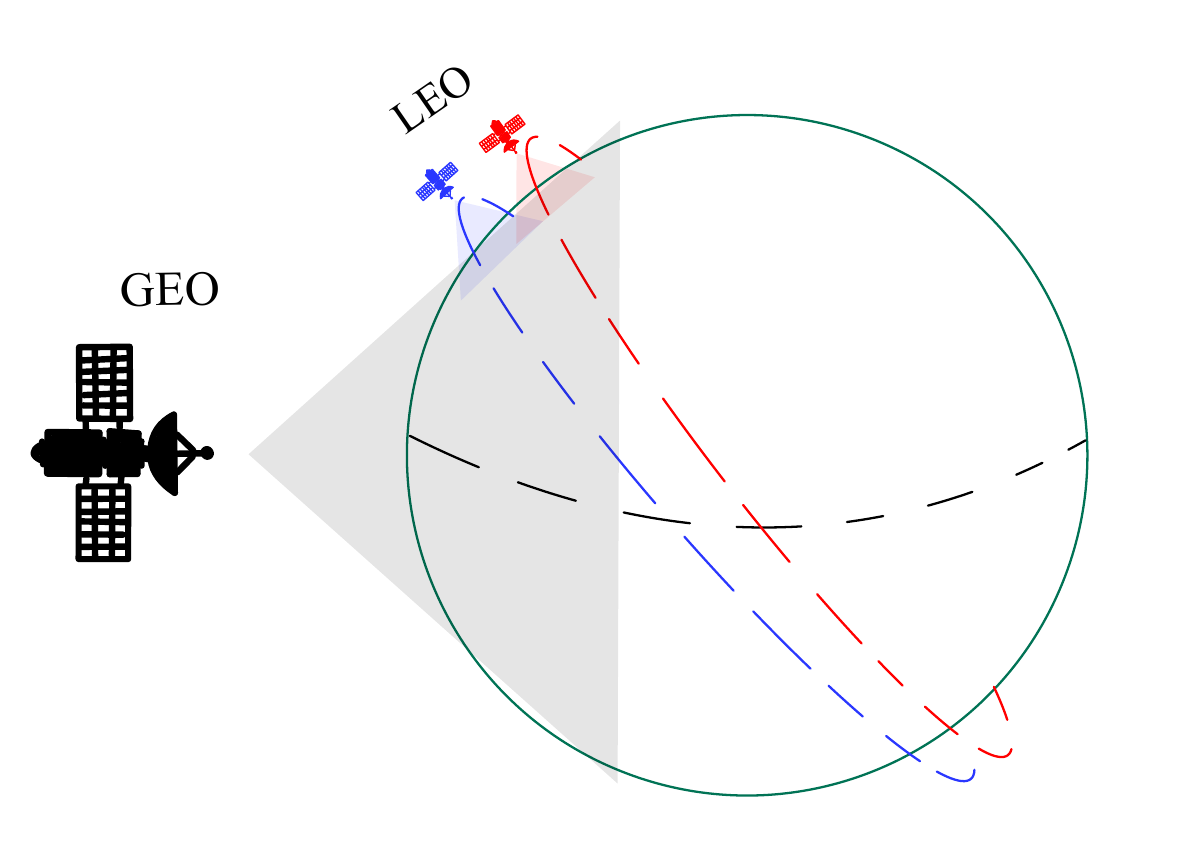}
    \caption{\textbf{LEO vs GEO Satellites}---% 
    \textnormal{LEO satellites closer distance provides lower latency at the cost of less coverage and no longer being in geostationary orbit.}}
    \label{fig:leo_v_geo}
\end{wrapfigure}

LEO satellite networks are comprised of hundreds to thousands of satellites that orbit 180~to 1,300~miles from Earth. LEO satellites' closer distance provides significantly lower theoretical latency ($\approx$10~ms RTT), at the cost of reduced coverage and a non-geostationary orbit (Figure~\ref{fig:leo_v_geo}).
%\todo{advantage: more back-up/redudancy relative to GEO}
The non-geostationary orbit and closer distance causes LEO satellites to travel at tens of thousands of miles per hour, orbiting the earth every 90~minutes. LEO satellite mobility is unique relative to other mobile networks (e.g., cellular, drones), because distances are longer and velocities are higher. Further, the core infrastructure of the network is constantly in motion, but theoretically predictable.
%the network is composed of thousands of routers (satellites) that provide substantially more bandwidth (i.e., terabytes per second).

% \begin{figure}[t]
%   %\centering
%     \includegraphics[width=\linewidth]
    
    \begin{wrapfigure}{R}{0.5\textwidth}
  %\centering
    \includegraphics[width=\linewidth]
    {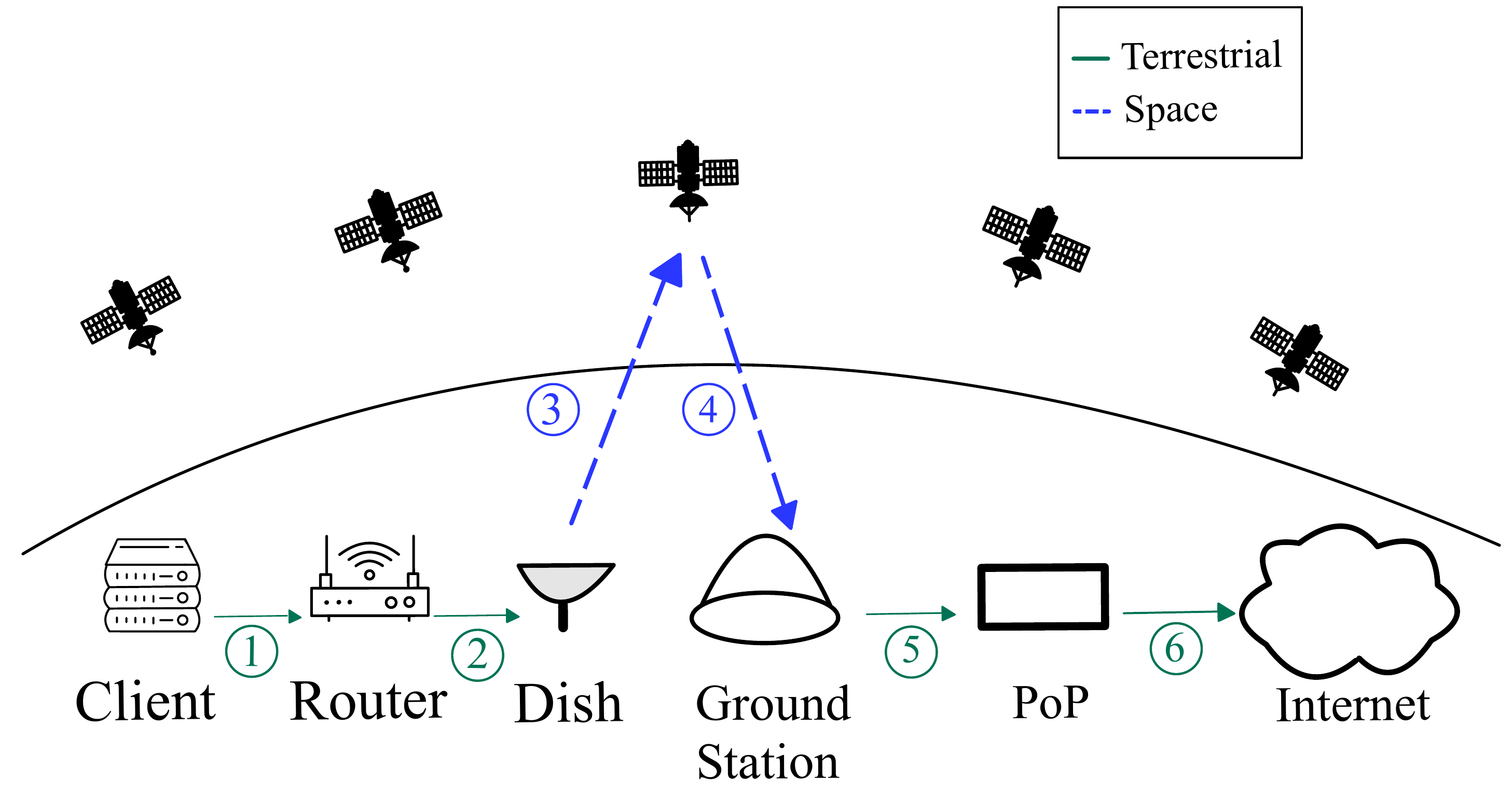}
    \caption{\textbf{Simplified LEO Satellite Network Routing (Egress)}---% 
    \textnormal{When a LEO client sends a packet to the public Internet, the packet is sent to a satellite using a dish, and is received by a ground station before forwarding to a POP. }}
    \label{fig:two_sat}
    \end{wrapfigure}
%\end{figure}

The most basic, and widely deployed~\cite{oneweb,tiansuan},  LEO architecture follow a ``bent pipe'' routing scheme (Figure~\ref{fig:two_sat}). Packet routing works as follows:
\begin{enumerate}
    \item A client sends a packet to its router
    \item The router forwards the packet to a physical dish
    \item The dish sends the packet via radio to a passing satellite
    \item The satellite relays the packet to a ground station
    \item The ground station forwards the packet to the provider's Point of Presence (POP), which is often located at an Internet Exchange Point and plays the equivalent role of a home gateway in mobile networks~\cite{mandalari2018experience}
    \item The packet is routed from the POP onto the Internet
\end{enumerate}

\noindent Some newer satellite architectures~\cite{starlink} are equipped with Inter-Satellite Links (ISLs), which allow satellites to relay packets to each other in space until a ground station is in view. ISL deployment provides coverage to clients in extreme remote locations (e.g., in oceans) that are in view of a satellite (e.g., within 600~miles), but not a ground station. ISLs send packets at the speed of light by using lasers in space, thereby surpassing the performance of optical fiber~\cite{chaudhry2022optical}.

% Other LEO architectures have evolved to solve the following accessibility problem: often, (e.g., in oceans), a satellite is in view of the client's dish, but not in view of a ground station. 
% Thus, 
% ISLs direct path of communication (unlike terrestrial fiber)~\cite{subcable_map}, and upwards of 100Gb/s~\cite{isl_bandwidth} bandwidth make them a favorable routing infrastructure.

% While a couple LEO satellite networks already provide Internet connectivity, many more \tcite{} are expected to roll-out in the coming years. 
As of 2023, Starlink~\cite{nytimes_star} remains the only consumer-targeted LEO satellite network. Starlink operates satellites with and without ISL capabilities~\cite{elon-lasers}. Other operational LEO satellites that target businesses include Oneweb~\cite{oneweb}, a network that caters towards enterprises that are located near the earth's poles (e.g., fishing companies in Alaska). Amazon's Project Kuiper~\cite{kuiper} and Telesat~\cite{telesat} are expected to deploy LEO satellites for consumer-targeted Internet in the future. 
% Starlink satellites provide coverage to populous areas and often do not reach near the earth's poles (e.g., Alaska)~\cite{hypatia}. 
% \todo{finish}
% \todo{Other operational LEO satellites exist, but ...}
% Oneweb~\cite{oneweb} is an operational LEO satellite that caters towards enterprises that are located near the earth's poles (e.g., fishing companies in Alaska).
% Oneweb satellite orbits provide more coverage to the earth's poles, compared to the most populous areas~\cite{hypatia}. 
% Oneweb satellites do not have ISLs.
% \todo{Telesat status? is oneweb the only other one that is operational?}
% Amazon's Project Kuiper ...
% Iridium uses LEO satellites to provide satellite telephony.\cite{iridium} 

%Satellite constellation orbital parameters and ground station locations can be found in FCC filings \tcite{\footnote{https://fcc.report/IBFS/Filing-List/SES }}.

\section{Related Work}

% The diversity of vantage points in LEO satellite research is uniquely important relative to GEO satellite measurement. 
% While the stationarity and large satellite-footprint of GEO satellites allowed researchers on the same continent to measure the same network connection~\cite{pavur2020tale,pavur2019secrets}, the mobility and small satellite-footprint of LEO satellites creates a dramatically different network for neighboring users~\cite{hypatia}.

Three primary methods have been introduced for studying LEO satellite networks: 
(1) buying and deploying one's own specialized hardware (i.e., a dish) to connect to satellite Internet, capturing real data but providing little coverage; 
(2) recruiting others with existing hardware, providing greater coverage at the cost of increased labor; or
(3) using theoretical models, providing the most coverage but no real data. 
Unfortunately, these methods face an inherent trade-off between collecting real data, coverage of vantage points, labor, and monetary cost. We detail each method below:

%\lhh minimizes all trade-offs by employing a key observation: exposed Internet services can reveal measurable LEO satellite routing. 

%\todo{shorten the next four paragraphs}

\vspace{3pt}
\noindent
\textbf{Deploying Physical Hardware.}\quad
To measure LEO satellite networks, researchers often buy and travel with specialized hardware (i.e., a satellite dish).  
Michel et~al.~\cite{imcdish22} deploy a satellite dish in Belgium to collect data about a single user's perceived Starlink latency. 
Ma et~al.~\cite{ma2022network} travel with four dishes around Canada to study Starlink latency across remote locations under different loads. 
Wang et~al.~\cite{tiansuan} deploy a constellation of satellites---Tiansuan Constellation---to act as an open research platform in space.

Relying on physical hardware presents a financial and coverage barrier to researchers:
hardware costs between \$500~\cite{starlink_cost}--\$23K~\cite{oneweb_cost}, while
%and over \$23K~\cite{oneweb_cost}; 
monthly subscriptions cost upwards of \$100.
To achieve world-wide coverage of LEO-accessible locations, one must travel with their hardware. 
%Traveling with hardware is complicated by LEO networks  waitlisting customers when arriving in new regions~\tcite{link.}
Furthermore, the majority of the world's population does not reside in a geographic location that qualifies for LEO satellite Internet subscriptions~\cite{starlink_coverage}. 

\vspace{3pt}
\noindent
\textbf{Recruiting Existing Hardware.}\quad
To collect data on LEO networks with a wider coverage of users and geographic locations, researchers often recruit existing LEO customers. 
Kassem et~al.~\cite{kassem2022browser} build browser extensions and recruit 18~Starlink users world-wide to study browser performance under Starlink connectivity.
RIPE Atlas~\cite{ripe_atlas}
has sent hardware (i.e., probes) to over 58 Starlink users across 13~countries, which can be used to run a variety of prescribed measurements.

Recruiting participants and maintaining data collection mechanisms (e.g., building extensions, sending hardware) creates a labor-consuming bottleneck for researchers and  produces only limited geographic coverage.
Further, relying solely on existing data collecting mechanisms often does not provide the data researchers need.
For example, RIPE Atlas measurements can only be conducted at minimum intervals of 60~seconds, which is too coarse to detect significant network fluctuations (Section~\ref{sub:sec:groundstations}).
% \todo{where to put}
% The Tiansuan constellation cannot be used to measure real-world user behavior or commercial LEO networks. 

% Using this to introduce inside-out vs outside-in
% While these efforts present an ``inside-out'' measurement of the network i.e., measurements from a vantage point inside the network, our methodology allows for on a ``outside-in'' measurement i.e., measurements from a vantage point outside to hosts inside the LEO satellite network thereby making measurements more accessible to researchers in all geographic locations.

\vspace{3pt}
\noindent
\textbf{Theoretical Models.}\quad
Given the difficulties in collecting empirical data, one of the most popular methodologies for studying LEO satellite networks is to use theoretical physics-based models, which simulate LEO Internet performance across location, satellite orbiting pattern, and congestion level.  
The most popular LEO simulators include Hypatia~\cite{hypatia}, Starlink.sx~\cite{starlinksx}, and SatelliteMap.space~\cite{satmap_space}.
The flexibility of these tools---and theoretical physics in general---has allowed researchers to simulate how LEO networks are affected by different congestion control algorithms~\cite{hypatia}, DDoS attacks~\cite{giuliari2021icarus}, route variability~\cite{bhosale2023characterization,wang2023reliability}, ISL deployment~\cite{bhattacharjee2023laser}, etc. 

However, the accuracy of most popular simulations~\cite{hypatia,starlinksx,satmap_space} has never been verified. 
The most recent LEO simulator published in NSDI 2023, StarryNet~\cite{lai2023starrynet}, does not evaluate latency predictions beyond the 90th percentile latency and is 20~times less accurate at predicting 90th percentile latency compared to 70th percentile latency.
Notably, our work shows that simulations are non-trivial to configure and do not always accurately model the architecture and the satellite--ground station selection process of 
%today's are selected in 
the deployed LEO satellite networks (Section~\ref{sub:sec:hypatia_eval}).  
%While simulations will always provide the most flexibility with regards to experiment configuration, our proposed method of \lhh provides substantially more coverage and experiment-flexibility relative to other existing data-collection solution. 

% \vspace{3pt}
% \noindent
% \textbf{Exposed Services.}\quad
% \lhh relies on exposed satellite services to measure satellite networks, but is not the first to use the presence and misconfiguration of devices online to measure Internet behavior. 
% For example, researchers used exposed services to study the impact of natural disasters~\cite{durumeric2013zmap}, track software patching behavior~\cite{durumeric2014matter}, identify spyware infrastructure~\cite{marczak2021candiru}, uncover clients of cyber espionage~\cite{marczak2020running}, unearth weaknesses in the Linux RNG~\cite{heninger2012mining}, and measure censorship~\cite{pearce2017augur, pearce2017global}.

\vspace{3pt}
\noindent
\textbf{Measuring Links.}\quad
We apply a decades-old observation---network characteristics can be remotely measured---to a new domain: LEO satellites. 
For example, in 1993 and 2006, Traceroute~\cite{malkin1993traceroute} and Paris-Traceroute~\cite{augustin2006avoiding} showed how routers could illuminate an arbitrary network path.  
In 1999, Downey et~al. introduced PathChar, a methodology for estimating latency and bandwidth between two arbitrary router hops~\cite{downey1999using}.
The same year Savage introduced Sting~\cite{savage1999sting}, a system for calculating packet loss rates across asymmetric routes.
Notably, LEO satellite routing introduces a new source of non-deterministic delay (i.e., moving satellites) that existing tools primarily attribute to (stationary) router queuing-delay.   
Throughout our work, we illustrate the new challenges and implications that LEO satellites surface when measuring network characteristics.

%Not sure if this adds to our story, lol 
% Our methodology leverages public-Internet exposed services hosted by LEO Internet providers. 
% While we are the first work to leverage the public-exposed services for network measurement, attackers have previously noticed the exposure of LEO-hosted user services on the public Internet. 
% Most recently, on October 11, 2022, Pro-Ukranian hacking group ``Team OneFist'' \emph{claims} to have successfully breached Russian LEO Satellites and compromised the configurations of multiple NovelSat NS3000 ground station modems in Moscow~\cite{attacktweet} through the use of an exposed service.

\vspace{3pt}
\noindent
\textbf{Inter-Satellite Links.}\quad
Prior work has perceived ISL's minimal latency as an opportunity to decrease latency and provide more direct routing paths~\cite{bhattacherjee2019network,bhattacherjee2018gearing,bhattacherjee2020orbit}, and has assumed Starlink does the same~\cite{hypatia,michel2022first}.
Proposed LEO attacks~\cite{giuliari2021icarus} and routing recommendations~\cite{bhosale2023characterization,wang2023reliability} also assumed ISLs correlate with low latency and direct routing.
However, we will show that prior work has overlooked a stark reality: while ISLs likely do minimize user to ground station routing, they significantly increase the length of the routing path. 
% between ground station to POP. 
Thus, customers who rely on ISLs may experience some of the highest latency and most indirect routing paths.
\section{LEO HitchHiking Overview}
\label{sec:hitchhike}

In this section, we present an overview of LEO \lhh, a general methodology to measure LEO satellite network characteristics at scale. \lhh builds on the key insight that probing publicly exposed satellite-routed devices 
can reveal the underlying satellite network architecture and performance characteristics. In contrast to previously used ``inside-out'' methodologies (i.e., connecting a measurement instrument to a satellite dish), which require physical access to privileged vantage points, the ``outside-in'' \lhh methodology requires no specialized hardware or painstaking recruitment.
\lhh can measure wherever satellite clients are already located across the globe.

Broadly, \lhh consists of three steps: 
(1)~identify publicly accessible endpoints (e.g., servers, routers) that transit LEO satellites for connectivity;
(2)~isolate where in the network path LEO satellites are used; and finally
(3)~craft an experiment to measure a desired characteristic (e.g., latency or availability) of the satellite link. 
In this section, we describe the general methodology. Then, in Section~\ref{sec:continuous}, we describe an implementation of \lhh specific to the Starlink LEO network and how it can be used to measure LEO latency.

\subsection{Finding Satellite-Routed Endpoints}

In the first step, \lhh needs to identify publicly reachable endpoints that transit a LEO satellite link
%To measure LEO satellite networks, \lhh needs reachable endpoints (e.g., servers, routers) that traverse a LEO %satellite link 
and are, ideally, geographically distributed. 
%Customer exposed services hosted by LEO networks can be used as measurable endpoints.
% \vspace{3pt}
% \noindent
% \textbf{Identifying LEO Networks.}\quad
To that end, \lhh must first identify networks (e.g., autonomous systems or IP blocks) that house LEO-routed services. 
Today, there exists only one commercial LEO network that sells to individual consumers: SpaceX-Starlink.
However, within coming years, AWS Kuiper, Telesat and OneWeb, are also expected to deploy consumer-oriented satellite services. 

% \vspace{3pt}
% \noindent
% \textbf{Filtering LEO-Hosted Services.}\quad
Once a network is identified, \lhh must find all Internet-exposed services that are hosted on the network.
Example services may include those that a customer wants to maintain remote access to, including  customer-exposed router administration portals or web servers. 
Notably, a service hosted within the address space of a LEO network does not immediately mean it transits a satellite link. 
For instance, LEO-network backbone equipment that routes traffic between a ground station and a POP, while within the LEO network's IP address space, does not necessarily traverse a satellite link.
Additionally, performance enhancement proxies (PEPs) are common in satellite routing as they decrease latency and increase reliability of networks by relying on proxies~\cite{pavur2021qpep}, caches~\cite{thibaud2018qoe}, or back-up non-LEO networks~\cite{pep}. While helpful to customers, PEPs add confounding factors to measuring LEO links. 
Thus, \lhh must filter for services that are likely using just a LEO-satellite for routing.

\subsection{Isolating Satellite Links}
\label{sub:sec:isolate_sats}
In the second step, after filtering for end-points that rely on LEO satellites, \lhh must identify satellite-based network hops. \lhh requires that enough of the routing path be visible to an external scanner, such that enough terrestrial routing artifact can be removed to make meaningful inferences. 
At a minimum, \lhh must 
(1) identify the hop before a satellite path is taken (i.e., hop~16 from Figure~\ref{fig:trace_pic}), and
(2) identify the hop after the satellite path is taken (i.e., hop~18 from Figure~\ref{fig:trace_pic}).
%\lhh can rely on traceroutes to identify hops that surround the LEO links in between. 

% It is possible that \lhh's external perspective might struggle to identify the immediate hop after the satellite path, as it will likely occur within an exposed service's local network (i.e., if the service's router transparently forwards packets).
% Without identifying the last hop, non-LEO artifacts (e.g., WiFi jitter) can add confounding factors. 
% Thus, to minimize non-LEO artifacts, \lhh must:
% (1) prioritize protocols that are most likely to elicit a router response (e.g., ICMP, ping), as opposed to another server's response (e.g., TCP, UDP) and%, to limit post-dish measurement to ethernet and 
% (2) apply a smoothing filter to results (Section~\ref{sec:continuous}) to eliminate short-lived artifacts.  
\subsection{Conducting An Experiment}
\label{sub:sec:isolate_experiments}

In the third step, having identified the satellite links in LEO satellite networks, \lhh can then be used to run measurement experiments.
%Notably, once the satellite link in the routing path is identified,
Many existing strategies for measuring characteristics of networking links (e.g.,~\cite{savage1999sting,downey1999using}) can be applied. 

%Crafting experiments with \lhh is simple. 
For example, to measure LEO satellite outages due to a customer's location or obstruction (e.g., a seagull lands on a dish), a researcher might do the following for all exposed LEO services in the same geographic area:
(1)~send a ping to the router before the satellite link (``terrestrial-hop'' router);
(2)~send a ping to the customer IP after the satellite link (``exposed-service'' router);
(3)~label potential outages as when exposed-service router pings are dropped, but terrestrial-hop router pings are not dropped, for an extended amount of time; and 
(4)~compare outages with neighboring exposed services, to determine if the outages are network-wide or customer-specific. 

%Many LEO other satellite measurements can be conducted using \lhh. For 
% As another example, to measure ISL usage, a researcher might calculate the geographic distance between the terrestrial-hop router and exposed-service router, to determine if the distance fits within coverage of a single satellite; if the distance does not fit within single satellite coverage, ISLs must be turned on. 
As another example, to measure LEO satellite bandwidth, a researcher might use the pathchar~\cite{downey1999using} approach, where packets of increasing size sent to both the hop before and after the satellite link can be used to estimate the satellite link's bandwidth without flooding the link.
%To minimize network artifacts from before the terrestrial-hop router impacting results derived from the LEO link, existing methodologies, such as increasing sample sizes~\cite{downey1999using}, can be applied. 
%To measure LEO satellite latency to determine if latencies match what network providers promise, a researcher might subtract the RTT of a ping of the terrestrial-hop router from the exposed-service router.

In this paper, we focus on measuring latency in the Starlink network.

% \subsection{\lhh Benefits}

% \lhh surfaces the following benefits:

% \vspace{3pt}
% \noindent
% \textbf{Collect real data.}\quad
% Networks rarely operate as hypothesized~\tcite{need examples!!}. 
% \lhh must collect data that is representative of how real LEO networks operate. 

% \vspace{3pt}
% \noindent
% \textbf{No specialized hardware.}\quad
% Specialized hardware is expensive~\cite{starlink_cost} and restrictive.
% \lhh must not require additional resources that an active Internet measurement researcher is not expected to have. 
% In other words, while active Internet measurement researchers are expected to have the compute and bandwidth to perform Internet scans, they are not expected to own satellite dishes or reside in a geographic location that qualifies for LEO satellite Internet subscriptions~\cite{starlink_coverage}. 

% \vspace{3pt}
% \noindent
% \textbf{World-wide view.}\quad
% LEO satellite Internet is expected to look vastly different depending upon the proximity of ground stations and satellite constellations~\cite{hypatia}. 

\subsection{Ethical Considerations}
\label{sub:sec:ethics}
\label{sub:sec:starlink_ethics}
\lhh relies on Internet-exposed services to measure LEO satellite links.
For the measurements in this paper, we follow the best practices outlined by Durumeric et~al.~\cite{durumeric2013zmap}, including configuring the scanner's IP to re-direct to an informative page that easily allows end-users to opt out of scans. We received no requests to opt out. 
We present \lhh to Starlink's engineering team and do not receive any reproach.

\lhh is not the first to use the presence of exposed devices to measure Internet behavior. For example, researchers have tracked software patching behavior~\cite{durumeric2014matter} and measured the impact of natural disasters~\cite{durumeric2013zmap} using exposed services. \lhh can also use application layer data, such as TLS certificates, to identify owners of a service. \lhh's use of application layer data remains consistent with the ethical standards followed by the community~\cite{lzr,akiwate2022retroactive,gigis2021seven}.

However, much like at the onset of Internet-wide scanning, we need to establish guardrails for the use of exposed satellite-based services for performing experiments. For example, LEO satellite links often operate at lower capacity than terrestrial links~\cite{michel2022first}. It is imperative that \lhh experiments do not degrade the quality of service for users by, for example, flooding LEO satellite links. 
We recommend researchers send the minimum number of packets needed to collect statistics about a LEO link, and avoid using tools that overload bandwidth (e.g., iperf~\cite{iperf}). 

%When scanning exposed services, researchers must never attempt to login to or access any non-public data and disclose vulnerable (e.g., leaking private information) exposed access points to their owners.

%When available, \lhh uses already-performed scans (e.g., Censys~\cite{censys15}) to reduce the overall number of probes sent.  

%In our work we find multiple instances of sensitive data exposed by governments and other organizations. 
%We are in the process of disclosing exposed access points to their owners.
%While our search criteria can be used to uncover vulnerable devices, these criteria are not difficult to uncover (e.g., simply searching \texttt{starlink} on Censys uncovers services). 

% \subsection{Summary}

% \lhh is the first data-driven methodology to measure LEO satellite networks that does not require physical hardware or recruiting participants.
% \lhh works in three steps:
% (1) find customer-exposed services in LEO networks,
% (2) identify the satellite link, and
% (3) define a measurable experiment on the satellite link. 
% In the next section, we describe an example system that uses \lhh to measure latency in the Starlink network.

\section{HitchHiking Starlink to Measure Latency}
\label{sec:continuous}

In this section, we present an implemention of \lhh to measure the latency of the only commercially available consumer-targeted LEO network: Starlink. 
Tracking network latency is particularly interesting in the LEO satellite setting, as LEO satellite mobility is expected to induce uniquely predictable and dynamic changes in routing paths~\cite{hypatia}. We measure latency by continually sending TTL limited pings on hitchhiked LEO links and collecting their round trip time.% and packet drop rate. 

We run a daily automated \lhh pipeline to measure latency with the following steps: 

\vspace{3pt}
\noindent
\textbf{1. Collect Exposed Services}\quad
To find measurable LEO-hosted Starlink services, we collect IPv4 and IPv6 exposed services in the Starlink network (AS\,14593) using Censys~\cite{censys15}, a public Internet device search engine. We note that researchers could also perform their own scans using tools like ZMap~\cite{durumeric2013zmap}, Masscan~\cite{graham2014masscan}, LZR~\cite{lzr}, and GPS~\cite{gps}. 

%\footnote{To find Internet-exposed satellite endpoints we use Censys~\cite{censys15}, a public Internet device search engine. 
%Censys performs continual scans of the full IPv4 address space and known IPv6 addresses (e.g., IPv6 address that appear in the Certificate Transparency Logs) on approximately 3,500~TCP and UDP ports. 
%Censys then attempts to detect L7 protocol and complete an L7 handshake, when possible.
%We note that  Censys does not currently scan any satellite-specific ports or protocols and may miss equipment that runs proprietary configurations. 
%Relying on Censys data reduces our scanning footprint. Exposed services inhabit hundreds of unexpected ports that would take non-negligible bandwidth to scan~\cite{}. \lhh also relies on Censys TLS handshakes to identify performance enhancement proxies and Censys DNS queries to identify customer IP addresses}.

\vspace{3pt}
\noindent
\textbf{2. Filter for Customer Endpoints.}\quad
To measure only services that likely use a LEO-satellite for routing, we filter for services that belong to customer endpoints. 
To measure Starlink customer services, we include only IP addresses whose DNS PTR record follow the Starlink customer format: \texttt{customer.\allowbreak[location].pop.starlinkisp.net}. There exist thousands of customer-exposed services in Starlink.
For example, on May 10, 2023,\footnote{All experiments in this section use data collected on May 10, 2023.} we identify a total 4,521~exposed services across 2,051~unique IPs (hosts), 857~unique ports, and 47~application layer protocols in the Starlink network.
After filtering for customer endpoints, 1,790~unique IPs remain.

\vspace{3pt}
\noindent
\textbf{3. Exclude PEPs.}\quad
We filter for performance enhancement proxies by automatically filtering IPs hosting a TLS certificate that belong to the most popular PEP within the Starlink network: Peplink, a PEP that combines 5G connectivity with Starlink.
Filtering for Peplink removes 9\% of all services (1,629 IPs remain). 
We manually analyze other TLS certificates and exposed services and do not find other identifiable PEPs. 

\vspace{3pt}
\noindent
\textbf{4. Geolocate Services.}\quad
To obtain an approximate geographic location of a Starlink service, we use Starlink's IP Geolocation feed~\cite{starlink-loc}.
Additionally, to identify the location of the customer's assigned POP, we (1) query the PTR record (e.g., \texttt{customer.atlagax1.pop.star\-linkisp.net}) associated with each host IP, (2) use the geographic location specified in the domain name (e.g., atlagax) to map the POP to a geographic location (e.g., Atlanta, Georgia)~\cite{reddit_mapping}. 
Often, the geolocation of a customer is not the same as the geolocation of a customer's POP, since most cities do not have a Starlink POP. 

% \begin{table}[t]
% %\centering
% \small
% \centering
% \begin{tabular}{llll}
% \toprule
% Hop & Router IP & RTT & Network \\
% & & (ms) & \\
% \midrule
% %10 & Continued... & ... & ... \\
% 11 & 2620:107:4000:8000::43 & 8  & Cogent \\ 
% 12-14 & \emph{No Response} & * & * \\
% 15 & 2620:134:b0ff::37d & 8 & Starlink \\ 
% 16 & 2620:134:b0fe:251::115 & 8 & Starlink  \\
% 17 & \emph{No Response} & * & * \\
% 18 & 2605:59c8:3049:fa00:44:facc:f696:eb2e & 45 & Dish \\
% \bottomrule
% \end{tabular}
% \vspace{8pt}
% \caption{\textbf{Truncated Traceroute From Public Server to Starlink Dish}---%
% \textnormal{ 
% The LEO link is traversed between the last hop and the second-to-last responsive hop, as indicated by the single spike in latency. }}
% \label{fig:alg_2}
% \vspace{-10pt}
% \end{table}

\begin{figure*}[t]
  %\centering
    \includegraphics[width=\linewidth]
    % {figures/sample_traceroute.pdf}
    {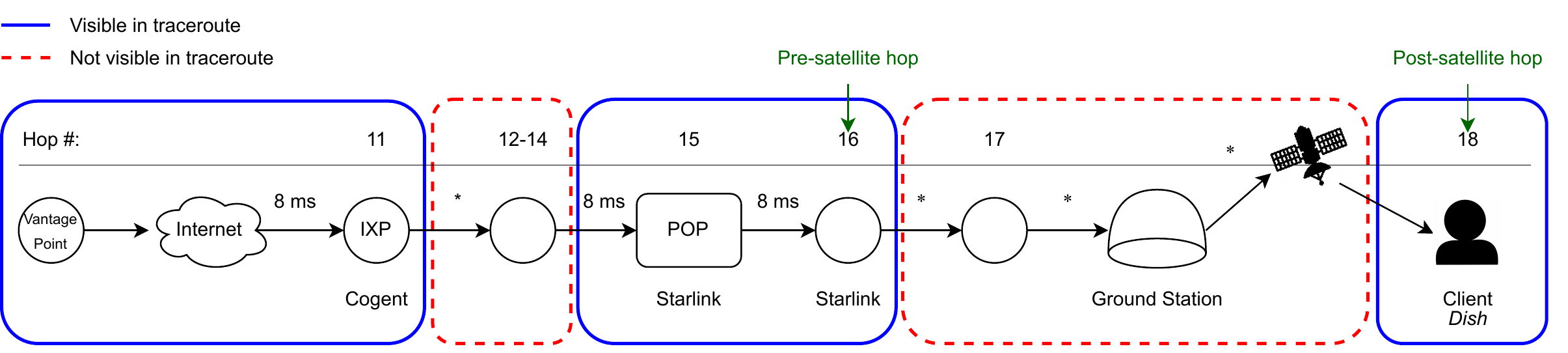}
    \caption{\textbf{Truncated ``Outside-In'' Traceroute From Public Server to Starlink Dish}---% 
    \textnormal{The LEO link is traversed between the last hop and the second-to-last responsive hop, as indicated by a jump in latency.}}
    \label{fig:trace_pic}
\end{figure*}

\vspace{3pt}
\noindent
\textbf{5. Identify Last Visible Pre-Satellite Hop.}\quad
We measure the satellite link using an ``outside-in'' perspective. 
In Figure~\ref{fig:trace_pic}, we show the output of running a traceroute\footnote{We experiment with many different traceroute tools in Appendix~\ref{app:traceroute}.} from \Stanford to an exposed Starlink service located in \SD that we control.\footnote{To connect to our dish from an external server, we
(1) configure the Starlink router to run on ``bypass'' mode,
(2) connect our own router, 
(3) configure our router to respond to pings, (4) identify the public IP address Starlink has assigned to us, and 
(5) ping our Starlink IP address (i.e., our router) from the external server.}.
The outside-in traceroute identifies the second-to-last hop (Hop~16) as the occurring right before the satellite link. 
Note, the hop immediately before the exposed service (Hop~17) is not visible.
In Appendix~\ref{app:sub:map_internal}, we use an internal network perspective to show that there is negligible RTT difference between the routers responsible for Hop~16 and~17. 

%, we show in Appendix~\ref{app:sub:map_internal}, using an internal network perspective how the non-responsive RTT is nearly equivalent to the previous hop's.  

\vspace{3pt}
\noindent
\textbf{6. Identify First Visible Post-Satellite Hop.}\quad
The last responsive hop in the traceroute is typically the first, and only, visible post-satellite hop. 
Notably, while the link between the last visible pre-satellite hop and the first visible post-satellite hop includes the satellite link, it also includes the terrestrial pathway between the POP and the ground station\@. 
In Section~\ref{sec:eval}, we show how our measurement technique filters much of the terrestrial routing outside of Starlink's network.
%as possible when compared to ground truth measurements. 

\vspace{3pt}
\noindent
\textbf{7. Measure Path Latency.}\quad
We measure path latency for 5~minutes sending two ICMP pings every second with the following additional configuration: (1) with a TTL equal to the terrestrial-hop-router-hop number and (2) with a TTL equal to the exposed-service hop number.
Notably, only one probe traverses the satellite link during each measurement, thereby minimizing ethical concerns. 
In Appendix~\ref{app:traceroute} and Appendix~\ref{app:manda_ping_details}, we show how sending TTL-specific pings increases coverage by an order of magnitude compared to TCP, UDP, and non-TTL specific pings. 

\vspace{3pt}
\noindent
\textbf{8. Isolate Satellite Link Latency.}\quad
We subtract the terrestrial-hop router RTT from the exposed-service router RTT, to measure the satellite link and minimize terrestrial artifacts.

\vspace{3pt}
\noindent
\textbf{9. Filter for Non-LEO Satellite Artifacts.}\quad
We apply a smoothing filter with a window size of 15~seconds (the time-step with which Starlink dishes stay connected to a satellite, before determining whether to switch connections~\cite{starlink_fcc}) to our timeseries of collected measurements, to eliminate short-lived artifacts (Section~\ref{sub:sec:lv_v_gt}). 
%Notably, \lhh might not always identify the immediate hop after the satellite path, as it will likely occur within an exposed service's local network (i.e., if the service's router transparently forwards packets).
Fortuitously, routers (one of the most popular hosts of exposed services) must be physically connected to a satellite dish using Ethernet~\cite{starlink_FOV}, thereby additionally minimizing potential Wi-Fi artifact. At least one-third of customer-exposed Starlink services belong to a vendor that produces routers and firewall appliances, with the two most poular being Fortinet and Sonicwall.

% \vspace{3pt}
% \noindent
% \textbf{10. Datastore.}\quad
% \lhh uploads the filtered measured satellite link latency to BigQuery, a scalable and serverless database platform, which we use for analysis.

%We discuss broader ethical considerations in Section~\ref{sub:sec:ethics}.

% we find that Starlink-exposed services help alleviate some ethical concerns.  
% We discover that Starlink-exposed services most likely belong to customers with higher bandwidth, minimizing \lhh's impact on end-users. The vast majority (95\%) of exposed IPs are IPv4 addresses, which most likely belong to enterprise customers (Appendix~\ref{app:how_expose}). 
% We manually investigate TLS certificates and identify 23~hosts that belong to a company (e.g., a prominent cruise ship company).
% %For example, two companies are in the maritime industry---an industry that makes popular use of satellite Internet~\cite{pavur2020tale}---including a prominent cruise ship company that exposed multiple login pages to Fortigate firewalls. 
% Enterprise customers are promised higher speeds, bandwidth, and reliability~\cite{starlink_business}.

\vspace{3pt}
\noindent
\textbf{10. Validating Incomplete Visibility}\quad
While \lhh lowers the barrier for identifying LEO satellite routing in the wild, it does not have complete visibility of all satellite routing. 
When measuring Starlink, \lhh cannot identify exactly what routing occurs between the POP and the client, does not know how many satellites, which satellites, and which ground stations packets are routed through.
To understand if the incomplete visibility is still sufficient, in the next section, we validate \lhh with physical equipment.
Critically, we find that \lhh's lack of visibility is not a limitation of \lhh; even Starlink customers with physical equipment have near identical visibility into Starlink's routing. 
In spite of this incomplete visibility, in Section~\ref{sec:www}, we demonstrate how \lhh's global perspective helps build informed inferences that illuminate previously undisclosed routing patterns. 

%\vspace{3pt}
%\noindent
Overall, the \lhh pipeline is designed to be quickly adaptable to new users and geographic locations. 
Since LEO network architecture changes nearly everyday due to new satellite launches~\cite{starlink_launches}, 
satellite falls~\cite{wsj_falls}, 
the integration of inter-satelllite lasers~\cite{spacex_lasers}, and new ground stations~\cite{starlink_more_ground_stations}, adaptability is crucial. 
The \lhh pipeline, which is open sourced under the Apache 2.0 licence, along with the data it collects, can be found at \url{https://github.com/stanford-esrg/LEO_HitchHiking}. 
%\url{[redacted-for-submission]}.
\section{Evaluation}
\label{sec:eval}

In this section, we evaluate the accuracy and coverage of \lhh on the Starlink network. 
First, we show that \lhh accurately measures satellite link latency relative to ground truth, capturing 100\% of all sustained latency spikes. 
Second, we compare \lhh's accuracy with the most popular LEO-network simulator and show that \lhh is up to 80\% more accurate. 
Third, we demonstrate \lhh's expansive coverage of LEO satellite links, which spans 13~countries and contains 28~times more measurable links than other methods.

% We first show that \lhh collects nearly identical latency data to ground truth as measured using a physical Starlink dish. 
% We compare \lhh measurements across three axes:
% (1) How close is the \lhh reported RTT to ground truth?
% (2) How often does \lhh capture the correct RTT spikes (i.e., true positive rate)?
% (3) How often does \lhh capture incorrect RTT spikes (i.e., false positive rate)? 
% Further, we compare, relative to ground truth, the accuracy and runtime (when applicable) of the two most popular alternatives to studying LEO network latency without deploying satellite equipment: Hypatia~\cite{hypatia} (the only available programmable  theoretical simulation of LEO networks) and RIPE Atlas~\cite{ripe_atlas} (one of the largest community Internet measurement platforms). 

\begin{figure*}[t]
  \centering
  \begin{subfigure}{\textwidth}
  \centerline{\includegraphics[width=\linewidth]{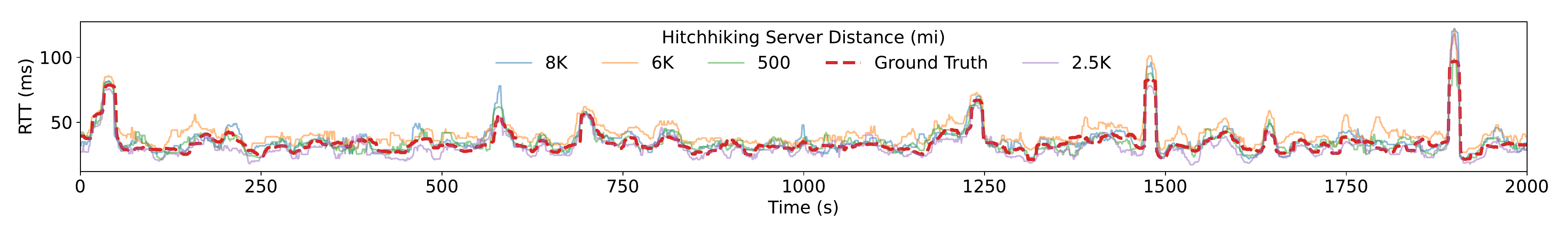}}
  \caption{Hitchhiking, Subtracting terrestrial-router hop}
  \label{fig:hitch_2hop}
  \end{subfigure}
  \hfill
    \begin{subfigure}{\textwidth}
    \centerline{\includegraphics[width=\linewidth]{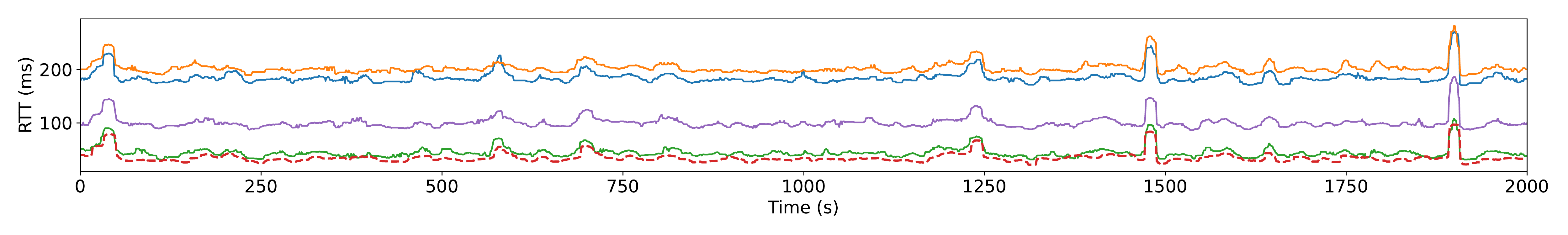}}
  \caption{Hitchhiking, Without Subtracting terrestrial-router hop}
  \label{fig:hitch_raw}
  \end{subfigure}
  \caption{\textbf{Comparing Hitchhiking With Ground Truth}---%
  \textnormal{\lhh from different servers world-wide is always able to detect long-lasting spikes in latency.
  %\todo{need bw compatible} 
  }}%}
\end{figure*}

\subsection{Comparison with Ground Truth} 
\label{sub:sec:lv_v_gt}

%We further compare the metadata that physical hardware provides relative to \lv. 

% \vspace{3pt}
% \noindent
% \textbf{Methodology.}\quad
To obtain ground truth about a client's LEO network latency, we deploy our own residential generation 2.0 Starlink\footnote{Starlink is the only commercially available LEO-provider that sells to individuals, and therefore the only network that easily provides ground truth.}
dish in \SD. 
To make our LEO link ``hitchhikeable,'' we 
(1) turn on router ping by bypassing the Starlink router with our own Asus RT-N66U router (Starlink generation 2.0 routers do not allow for port forwarding or respond to ICMP probes), and 
(2) configure our router to advertise the Starlink-provided public IPv6 address.

We also collect Starlink provided metrics from our dish every second, including the reported “POP ping latency,” packet drop, and estimated bandwidth usage. 
The POP ping latency is the ground truth RTT of a ping from our dish to the assigned POP\@. 
The POP ping latency is the most granular provided metric of LEO latency; it includes terrestrial latency to and from the ground station. 
Starlink metrics do not reveal which ground station or satellite the dish is connected to~\cite{nelsonslog}. 
However, we use a novel side-channel in our Starlink dish, the obstruction map, which in real time records the location of a successful satellite connection.
We describe in Appendix~\ref{app:obstruction_meth} our side-channel methodology. 
% Thus, to track satellite connections in real-time, we reboot our dish and save the obstruction map every second. 
%Figure~\ref{fig:sat_heatmap} and~\ref{fig:sat_change} shows this ground-truth data on satellite connections, and corresponding latency collected from the dish.

\subsubsection{\lhh}
%For our experiment, we hitchhike the dish's IPv6 address. % on May 12, 2023 
We compare \lhh measured latency to ground truth. 
To evaluate \lhh under different distances from \SD, we run \lhh from four geographic locations: Australia, Brazil, California (US), and Virgina (US).
Each location is 500--8,000~miles away
%\footnote{For anonymity, we do not reveal the exact distance of each location to our dish.}
from our dish. 
All \lhh pipelines run at the same time on May 12, 2023.
We note the presence of clear skies and no physical obstructions, and thus minimal interference during our experiment. 

\vspace{5pt}
\noindent
\textbf{Evaluation.}\quad
Across all \lhh vantage points, \lhh captures latency statistics that are close to the ground truth: 96\% of reported RTT times are within one standard deviation (10~ms), and 50\% of RTT times are within 3~ms of the ground truth (i.e., the dish's pop ping latency).
We illustrate the output of \lhh relative to the ground truth in Figure~\ref{fig:hitch_2hop}
and the output of \lhh without removing terrestrial artifact---to better depict the captured RTT patterns across \lhh locations---in Figure~\ref{fig:hitch_raw}. 
\lhh captures 100\% of all sustained RTT spikes, which we define as latencies over two standard deviations away from the median that last at least 15~seconds (i.e., the Starlink minimum amount of time dishes stay connected to the same satellite~\cite{starlink_fcc}). 
We describe the underlying cause for sustained RTT spikes (ISL usage) in Section~\ref{sec:www}. 
%https://web.archive.org/web/20220320174537/https://ecfsapi.fcc.gov/file/1020316268311/Starlink%20Services%20LLC%20Application%20for%20ETC%20Designation.pdf

\lhh observes only one false-positive RTT spikes across our cumulative 10,000 second measurement period.
All but one \lhh geographic location has a 0\% false positive rate for capturing RTT spikes. 
When \lhh is deployed 6,000~miles from our ground-truth dish, it returns one false positive RTT peak near second~1800. 
We attribute the false positive to a less stable terrestrial-router hop; while other terrestrial-router hops never deviate more than 1~ms from the average, the 6,000~miles terrestrial-router hop jitters up to 10~ms in latency near second 1800.
Thus, although terrestrial-router RTT is subtracted from the final hop, jitter can propagate.
%When analyzing \lhh RTT spikes, to increase result confidence, users can filter for endpoints in which terrestrial-router hops never deviate more than 1~ms in latency.
To decrease the false positive rate of finding sustained latency spikes, in Section~\ref{sec:www}, we only study endpoints whose second-to-last hop experiences no more than 1~ms deviation of RTT.

% \section{Filtering Root Causes For Starlink Latency Dynamics}
% \label{app:rtt_dish_causes}

\begin{figure*}[t!]
  \centering
    \begin{subfigure}[t]{0.45\linewidth}
  \centering
  \includegraphics[width=\linewidth]{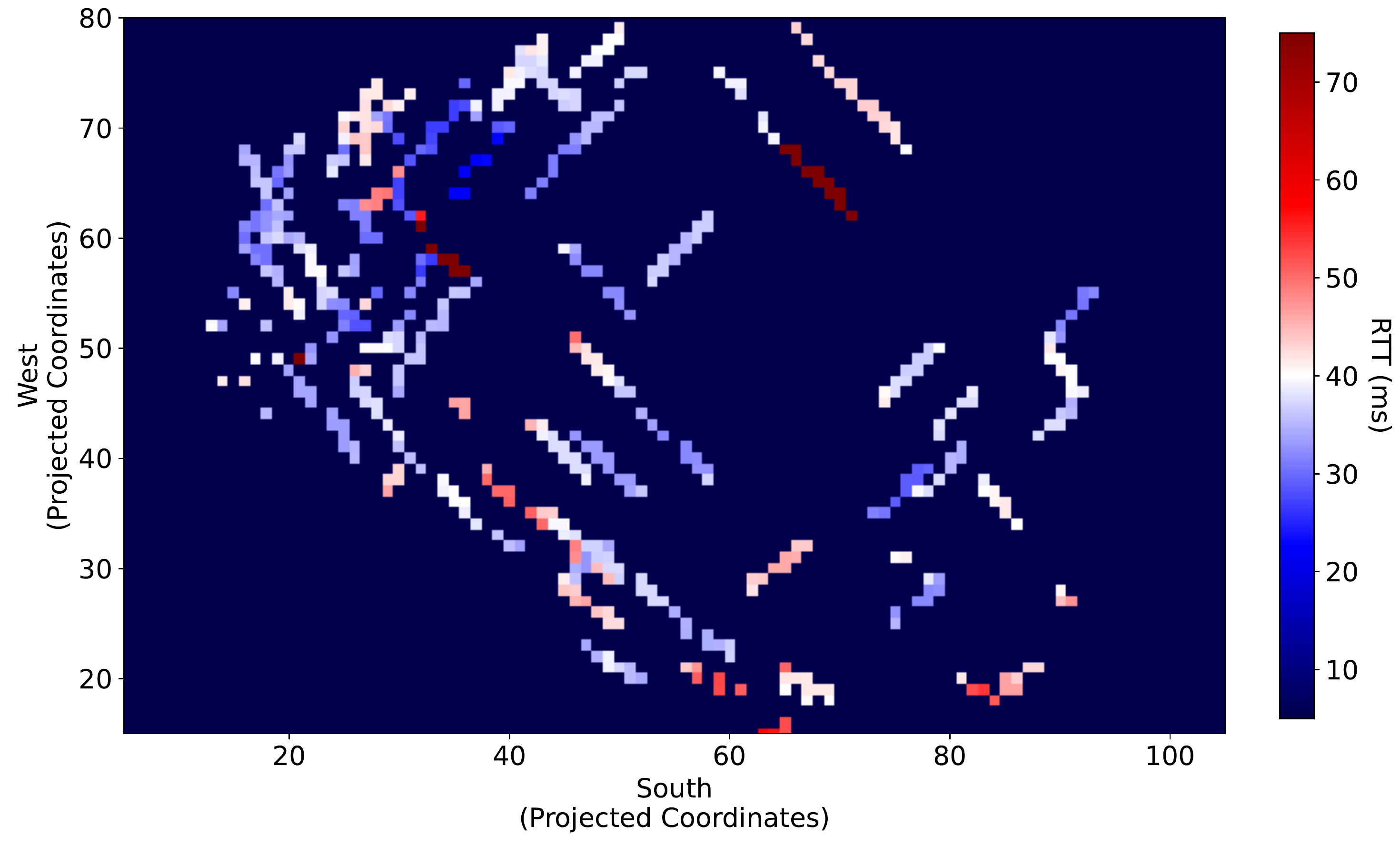}
  \caption{\textbf{Sustained Latency Spikes are Not Correlated With Satellite Location}---\textnormal{An image projection of the satellites the dish connects to when facing the sky (Appendix~\ref{app:obstruction_meth}), overlayed by the user's RTT when connected to each satellite's location, shows no correlation between high RTTs and satellite location. }}
  \label{fig:sat_heatmap}
  \end{subfigure}\hfil
  \begin{subfigure}[t]{0.45\linewidth}
    \centering
    \includegraphics[width=\linewidth]{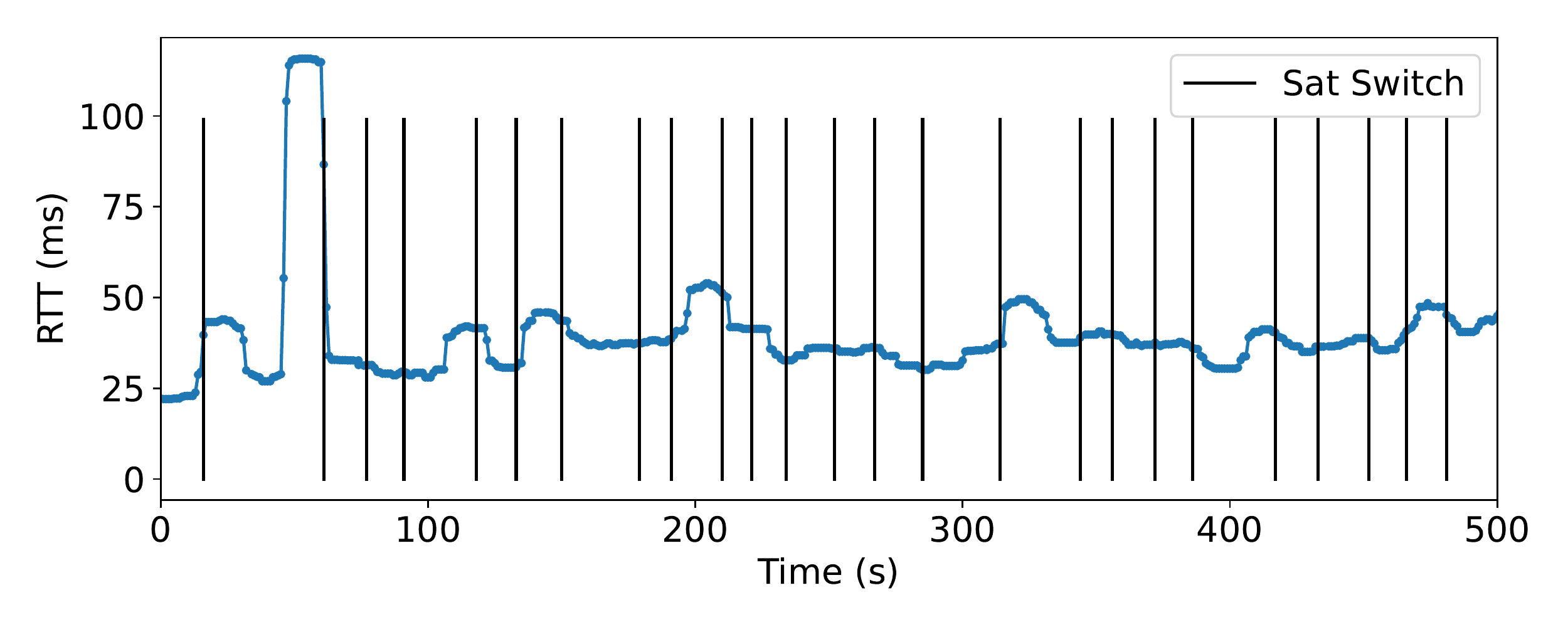}
  \caption{\textbf{Sustained Latency Spikes are Not Always Satellite Changes}---\textnormal{Satellite switches are computed using the obstruction map above. } }
  \label{fig:sat_change}
  \end{subfigure}
  \caption{\textbf{Satellite Location Relative to Latency}---\textnormal{Sustained increased latency is not due to satellite location or satellite changes.}}%}
\end{figure*}

\label{sub:sec:obstruction_map}
Sustained spikes in RTT are not caused by distant satellite location and not always caused by satellite switches. 
In Figure~\ref{fig:sat_heatmap}, we present the obstruction map overlayed with the corresponding ground truth RTT.
There exists no clear correlation between sustained latency and the satellite locations relative to the dish: sustained anomalous RTTs, colored in red, occur throughout all satellite locations. 
In Figure~\ref{fig:sat_change}, we vertically mark every satellite change, which we detected when the dish connects to a satellite that is not neighboring the prior connection's location.
Within the first 500 seconds of measurement, the first sustained RTT peak occurs while still connected to the same satellite.
Across the entire measurement period, 2/5 sustained RTT spikes occur while still connected to the same satellite.
For standard RTT spikes (i.e., spikes over one---but not two---standard deviations above the median), 5\% occur while still connected to the same satellites. 
Thus, satellites switches are not the ultimate cause behind latency spikes. 

RTT spikes are also not due to congestion. 
The ground truth metrics report  no packet drop or drop in bandwidth during sustained or standard latency spikes. 
Furthermore, we find that spikes occur in multiples of 15~second---aligning with Starlink's reported satellite reconfiguration period~\cite{starlink_fcc}---further making any cause of latency that is independent of Starlink routing (e.g., brief congestion caused by a nearby user) unlikely.  
In Section~\ref{sec:www}, we use \lhh's global perspective, and validation by Starlink, to show that sustained latency spikes are due to routing path changes exacerbated by ISLs.
Notably, Starlink shares that routing path changes can happen even if a user remains connected to the same satellite. 

% \lv also measures the latency of 17~customer exposed services in OneWeb. 
% In Table~\ref{tab:topLocations_one}, we show that OneWeb enterprise customers are located in northern territories---Alaska and Canada.
% The northern customers are consistent with OneWeb's business model in which satellite constellation deployment provides more coverage for the northern poles. 
% RIPE Atlas does not have any probes that use the OneWeb network, and is therefore unable to predict OneWeb latencies. 

\subsubsection{LEO Simulations}
\label{sub:sec:hypatia_eval}
\label{sub:sec:groundstations}
\label{sub:sub:sec:hypatia}

\begin{figure*}[t!]
  \centering
    \begin{subfigure}[t]{0.45\linewidth}
  \centering
  \includegraphics[width=\linewidth]{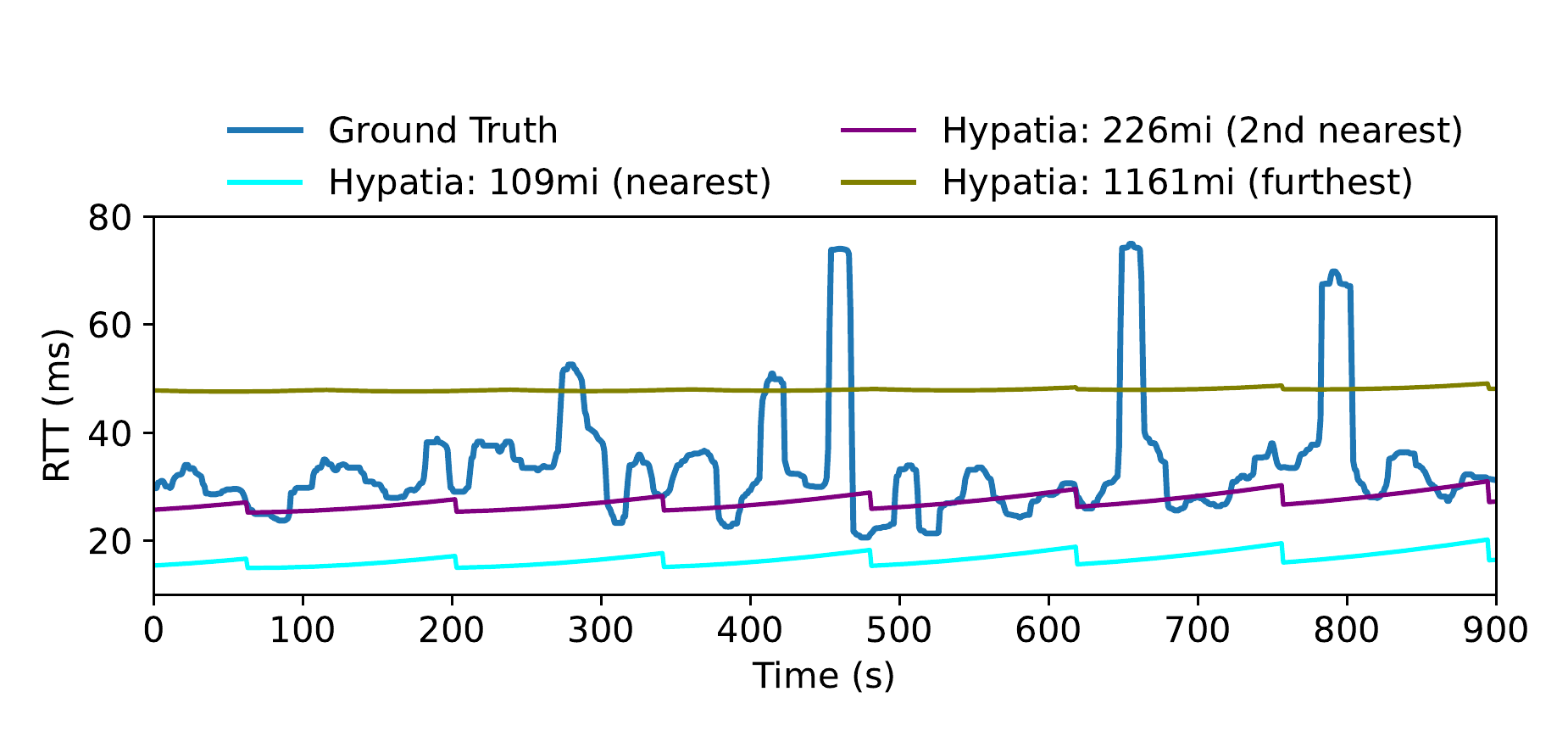}
    \caption{\textbf{Simulation Predictions}}
    \label{fig:best_hypatia_star}
  \end{subfigure}
  %\hfill
      \begin{subfigure}[t]{0.45\linewidth}
      \centering
   \includegraphics[width=\linewidth]{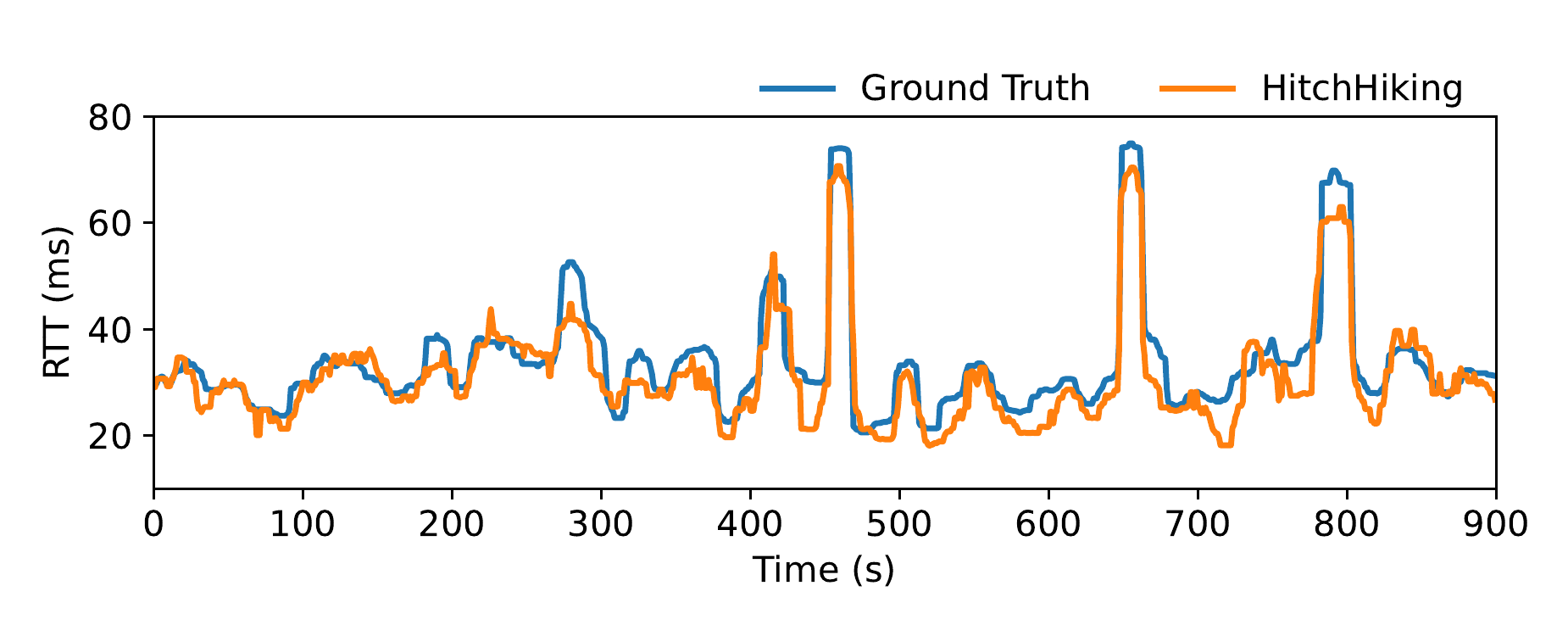}
  \caption{\textbf{\lhh}}
    \label{fig:star_v_lv}
  \end{subfigure}
      \caption{\textbf{\lhh Comparison With Prior Work}---
      \textnormal{Simulations do not capture the dynamics of real-world Starlink RTTs.
    }}
\end{figure*}

LEO simulations model real-world LEO networks to help researchers explore scenarios that are impossible to test on real networks. 
We consider two LEO simulators: (1) the most widely-used, Hypatia~\cite{hypatia} and (2) the newest, StarryNet~\cite{lai2023starrynet}.

Hypatia is a LEO simulator that takes as input the geographic location of ground stations and satellite constellation parameters (e.g., number of satellites, their altitude, inclination, etc.). Hypatia then returns the predicted RTT packet latency between two ground stations (GS) A and B. We show that \lhh is more accurate than Hypatia at estimating LEO latency. 
Unfortunately, we cannot evaluate against StarryNet because it requires over 2~TB of RAM to simulate Starlink and is not able to run in cloud environments\footnote{We inspect the source code and find that StarryNet requires a specific type of local network reconfiguration which Google Cloud cannot successfully execute.}. 
% Unfortunately, we cannot evaluate against StarryNet because it requires over 2~TB of RAM to simulate Starlink, and cannot successfully run on the Cloud\footnote{The authors of StarryNet deploy StarryNet on a cluster of local machines that give StarryNet access to reconfigure the local network during simulation. We do not run StarryNet on a local machine because it suspiciously requires the plaintext root credentials of its host machine as input, and suspiciously contacts a foreign server thousands of times during simulations. }. 

\vspace{3pt}
\noindent
\textbf{Methodology.}\quad
% from our dish and use it to scan our dish's exposed service. 
To model Starlink latency, we configure Hypatia with the Starlink constellation parameters published by the FCC~\cite{sat_earth_station_fcc}. 
To model client latency to a POP, we run multiple simulations where GS A is the location of our dish (Section~\ref{sub:sec:lv_v_gt}) and GS B is a GS within the set of all GS that are reachable using one satellite hop from the dish.
% We assume ISLs are not being used, as our dish is within one satellite hop to 9~ground stations and the POP itself.\footnote{In Section~\ref{sec:www}, we infer that clients near ground stations and POPs are likely using ISLs.
% However, we find that while ISL usage in Hypatia decreases RTT, 
% ISLs likely increase RTTs in real world deployment.Thus, if we were to configure Hypatia to use ISLs, Hypatia-estimated latency would decrease and cause it to be even less accurate. } 
To account for additional terrestrial latency between the ground station and POP, we add the latency that a packet would incur traveling at 2/3rds the speed of light (i.e., estimated optical fiber latency~\cite{chaudhry2022optical}) between the ground station and POP\@. 

We use Hypatia's default satellite selection algorithms: Hypatia connects the client with a satellite that minimizes the RTT between the client and ground station.
In Appendix~\ref{app:eval}, we present the results of configuring Hypatia using a theoretical worst case satellite selection algorithm and find that it does not significantly change the quality of Hypatia's predictions. 
%Since we compare only against ground stations that are within one satellite hop, Hypatia itself assumes that ISLs will never be used.
We compare Hypatia results with our ground truth dish and a \lhh deployment located 500~miles away from the dish.
\begin{wraptable}{R}{0.5\textwidth}
%\centering
\footnotesize
\centering
\begin{tabular}{llll}
\toprule
DNS  & POP & \multicolumn{2}{c}{\# Distinct IPs} \\ \cmidrule{3-4} 
Subdomain&  Location  & Hitch- & RIPE \\
&   & Hiking & Atlas \\
 
\midrule
\texttt{sttlwax1}  & Seattle, Washington & 243 &  7\\
\texttt{atlagax1}  & Atlanta, Georgia & 210 & 4 \\
\texttt{dllstxx1}  & Dallas, Texas & 186 & 1\\
\texttt{chcoilx1}  & Chicago, Illinois & 182	&6 \\
\texttt{lsancax1}  & Los Angeles, California & 173 & 2 \\ 
\texttt{sydyaus1}  & Sydney, Australia & 148&4	\\
\texttt{nwyynyx1}  & New York City, New York	 & 144& 6\\
\texttt{frntdeu1}  & Frankfurt, Germany & 124 &13\\
\texttt{dnvrcox1}  & Denver, Colorado & 87	& 7 \\
\texttt{lndngbr1}  & Heathrow, England & 56&	 6\\
\texttt{mdrdesp1}  & Madrid, Spain & 20& 0\\
\texttt{sntoch1}  & Santiago, Chile & 19& 1 \\
\texttt{acklnzl1}  & Auckland, New Zealand & 11& 0\\
\texttt{lgosnga1}  & Lagos, Nigeria & 6 &0\\
\texttt{bgtacol1}  & Bogata, Columbia & 5& 0\\
\texttt{limaper1}  & Lima, Peru & 3&0 \\
\texttt{prthaus1}  & Perth, Australia & 3&1\\
\texttt{qrtomex1}  & Mexico City, Mexico & 3 &0\\
\texttt{splobra1}  & San Paulo, Brazil & 3&0 \\
\texttt{tkyojpn1}  & Tokyo, Japan & 3& 0\\
\midrule
Total & & 1,629 & 58\\
\bottomrule
% Los Angeles, California 2 & 22	 \\ % not including because doesn't follow naming convention
% Madrid, Spain & 4	 \\
% Auckland, New Zealand & 4	 \\
% Santiago, Chile & 2	 \\
% Guarulhos, Brazil & 1 \\
% Mexico City, Mexico	& 1 \\
\end{tabular}
\vspace{8pt}
\caption{\textbf{Geographic Coverage of Exposed Services}---%
\textnormal{ 
Starlink exposed services are geographically wide-spread, providing \lhh an ample amount of measurable LEO links.
}}
\label{tab:topLocations_star}
%\vspace{15pt}
\end{wraptable}

% \begin{table}[t]
% %\centering
% \small
% \centering
% \begin{tabular}{llll}
% \toprule
% Customer & Customer  & \multicolumn{2}{c}{\# Distinct IPs} \\
%  & Location & LEO- & RIPE \\
%  & & Voyager & Atlas \\
%   \midrule
%  Pacific Dataport & Anchorage, Alaska & 15 & 0 \\
%  Northwestel & Yellowknife, CA & 2 & 0 \\

% \bottomrule
% \end{tabular}
% \vspace{8pt}
% \caption{\textbf{\lv Geographic Coverage of OneWeb}---%
% \textnormal{ 
% \todo{}.}}
% \label{tab:topLocations_one}
% %\vspace{-15pt}
% \end{table}

\vspace{3pt}
\noindent
\textbf{Evaluation.}\quad
While \lhh accurately estimates LEO satellite latency relative to ground truth, Hypatia is difficult to parameterize such that its output matches ground truth. 
In Figure~\ref{fig:best_hypatia_star}, we illustrate the output of Hypatia when using the best-case satellite selection algorithm, and a subset of nearby groundstations (the nearest, second nearest, and furthest). 
No matter the ground station, Hypatia never predicts that a client experiences sustained RTT spikes, unlike \lhh (Figure~\ref{fig:star_v_lv}).
In Section~\ref{sub:sub:sec:isl_prevale}, we find that RTT peaks are due to dynamic ISL routing patterns, which Hypatia is unaware of. 
Notably, configuring Hypatia to use the second nearest ground station produces a latency prediction that on average is only 7.6~ms in error relative to the ground truth.
Nevertheless, \lhh RTTs are on average 1.8~times more accurate than Hypatia.

Hypatia's inability to model real-world LEO links is not necessarily a deficiency of Hypatia, but rather a limitation of applying theoretical models to predict latencies about a network that reveals little about its operation. Starlink does not reveal its internal fiber paths between ground stations and POPs, terrestrial routing decisions, satellite selection algorithm, ISL routing, or congestion patterns.
In order to approximate latency, the only ground truth information Hypatia has access to is satellite and ground station location.
%user-selected ground stations and, optionally, arbitrary injections of congestion. 

% https://www.auroraiv.com/solutions/
% Talkeetna ground station

%\lhh is also an order of magnitude faster than Hypatia. 
%While simulations are tough to configure to reflect ground truth, they also have a prohibitive runtime relative to active measurement.
% While collecting a 30~second sample of LEO network latency consumes 30~seconds of \lhh (assuming an exposed service has already been found), simulating 30~seconds of RTT between just two ground stations consumes 1~hour of wall-clock time on a 12~core machine with 58~GB of RAM\@.
%A 6~hour Hypatia simulation executes in 41~hours.
%The length of simulated 
% As simulations grow longer, per-simulated-second wall-clock time decreases, but not substantially: 30~min simulations execute in 4.5~hours and a 6~hour simulation executes in 41~hours. 
%Thus, \lhh is up to 120~faster at estimating LEO latency than simulations. 

Nevetheless, Hypatia's theoretical minimum calculations can help illustrate which ground stations a dish is connected to. 
Given that 20\% of ground truth RTT fall below the second-nearest ground station best-case RTTs, Starlink must be connected to the nearest ground station at least 20\% of the time.
However, using Hypatia alone, it is unclear whether other increases in RTT latency are due to connecting to a different ground station, or congestion.

% \begin{figure}[t]
%   %\centering
%     \centerline{\includegraphics[width=\linewidth]{figures/oneweb_xtreme.pdf}}
%     \caption{\textbf{Simulated OneWeb Satellite Selection With Talkeetna Ground Station}---% 
%     \textnormal{\todo{  }}}
%     \label{fig:xtreme_hypatia_oneweb}
% \end{figure}

\subsubsection{RIPE Atlas}
RIPE Atlas---one of the most comprehensive community-driven Internet measurement platforms---provides substantially less coverage and granular statistics than \lhh. 
Unfortunately, the nearest RIPE Atlas dish that is assigned our POP is located over 300~miles away (i.e., uses a different set of ground stations). 
Furthermore, RIPE Atlas limits measurements to occur no more frequently than every 60~seconds, making it difficult to capture sustained RTT spikes that often change every 15~seconds.

% (e.g., latency spikes that occur for a consecutive 15~seconds).  
% Nevertheless, in Figure~\ref{fig:star_v_lv}, we plot a hypothetical best case RIPE Atlas measurement scenario, in which our ground truth is sampled every 60~seconds.
% RIPE Atlas' mean RTT is 33\% lower than the ground truth's and 4.5~times more inaccurate than \lhh.
% RIPE Atlas's coarse sampling also misses two out of three sustained RTT spikes. 

% We attempt, but are unable, to use RIPE Atlas to also approximate our dish's POP latency over time. 
% We first search for the RIPE Atlas probe closest to our dish to measure from. 

%RIPE Atlas does not have any probes that use the OneWeb network, and is therefore unable to predict OneWeb latencies. 

\subsection{\lhh Coverage}
\label{sub:sec:coverage}

\lhh provides the most comprehensive coverage of LEO networks today. 
Fortuitously for \lhh, LEO network customers expose services across the world. 
On May 10, 2023, we use \lhh to measure all (publicly exposed) LEO links in the Starlink network.
In Table~\ref{tab:topLocations_star}, we list the POP locations of Starlink customers that expose services, as well as the POP locations of all RIPE Atlas Starlink Probes.
\lhh has 28~times more exposed services than RIPE Atlas, whose services use POPs across only five countries.
Only seven IPs found by \lhh overlap with those of RIPE Atlas.
\lhh finds exposed Starlink services are assigned to POPs that reside across 20~cities and 13~countries.
While a majority (59\%) of \lhh found services use POPs in the US, there exists a long-tail of other locations including Australia (7\%), Germany (6\%), and England (3\%). 

%While LEO simulations can provide the most coverage, they do not provide real data. 
%In Section~\ref{sec:limitations}, we discuss the limitations of using POPs and enterprise locations to approximate user geolocation, and how future work can leverage RIPE Atlas to improve LEO geolocation accuracy. 

\section{Worldwide Latencies in the Wild}
\label{sec:www}

We perform the most geographically-diverse data-driven analysis of LEO satellite latency to date.
Our global perspective illuminates that real world deployment of a global LEO network is more complex than previously understood. 
While prior work attributed differences in customer latency to localized effects such as satellite location or congestion, we infer, and validate, that customer latency is correlated with a customer's distance to POP and unexpected ground station selection.
Additionally, our investigation surfaces an overlooked reality by prior work: while ISLs do increase coverage, they significantly increase the distance of the route between the ground station to POP. 
%Previously (Section~\ref{sub:sec:obstruction_map}), we found that congestion, satellite location, and satellite switches were unlikely to be the primary causes behind increases in latency. 
%\todo{concluding sentence}
%We measure users across farms, boats, and other remote locations 
% \begin{wrapfigure}{r}{0.5\textwidth}
% %\centering
% \includegraphics[width=\linewidth]{figures/best_rtt_ripe.pdf}
% \caption{\textbf{Latency vs POP Distance}---% 
% \textnormal{Customers located further from their assigned POP experience higher minimum RTT times. }}
% \label{fig:latency_v_pop}
% \end{wrapfigure}

We use the \lhh methodology (Section~\ref{sec:continuous}) to collect Starlink latency data between May~18--June~23, 2023.
Additionally, we use \lhh to scan the IP addresses of all Starlink RIPE Atlas probes, which provide ground truth about customer location; we mention when RIPE Atlas probes are used in our analysis. 
We filter the initial 3.5K exposed customer Starlink IPs for jittery pre-satellite hops, leaving 2.4K IPs that host exposed services.
The set of services follows a geographic distribution similar to Table~\ref{tab:topLocations_star}.
%We filter the initial 1.6K exposed customer Starlink IPs for jittery pre-satellite hops, leaving 989 IPs that host exposed services.

% One experiment requires ground truth about a customer's location relative to their POP.
% While exposed services do not always readily provide customer location, RIPE atlas probes do. 
% Thus, we additionally configure \lv to scan the IP addresses of all Starlink RIPE atlas probes.
% %, of which only eight are not filtered (i.e., do not belong to dual-homed set-ups, are reachable, have a terrestrial-router hop that is reachable). 
% We explicitly mention when \lv-measurements of RIPE atlas probes are used in our experiment. 

% Nigerian-POP customers experience the worst average RTT (110~ms) and standard deviation (86~ms).
% Standa
% Brazil-POP customers experience the second largest average RTT (104~ms) and standard deviation of RTT (27~ms).
% Peru, Australia, and New Zealend experience the shortest RTT times, on average under 30~ms. 

\begin{wrapfigure}{r}{0.5\textwidth}
%\centering
\includegraphics[width=\linewidth]{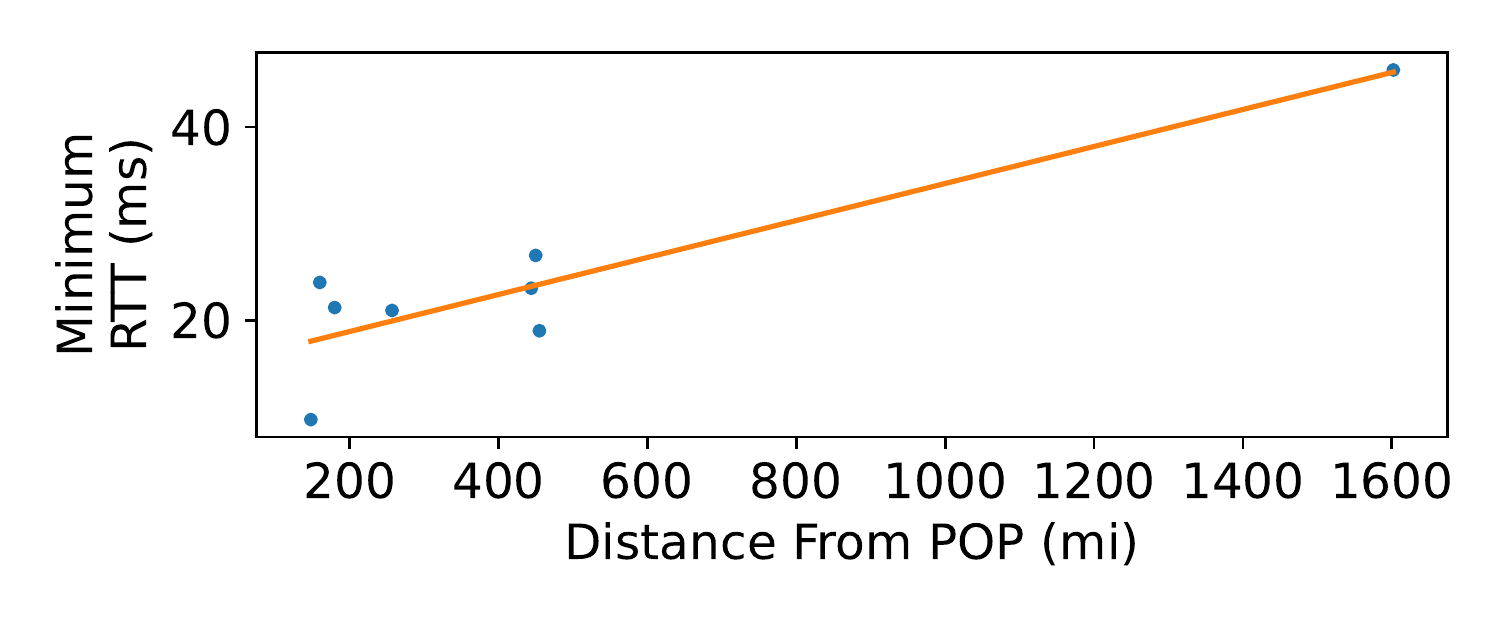}
\caption{\textbf{Latency vs POP Distance}---% 
\textnormal{Customers located further from their assigned POP experience higher minimum RTT times. }}
\label{fig:latency_v_pop}
\end{wrapfigure}

\begin{wrapfigure}{R}{0.5\textwidth}
  %\hfill
    \begin{subfigure}{0.5\textwidth}
    \includegraphics[width=\linewidth]{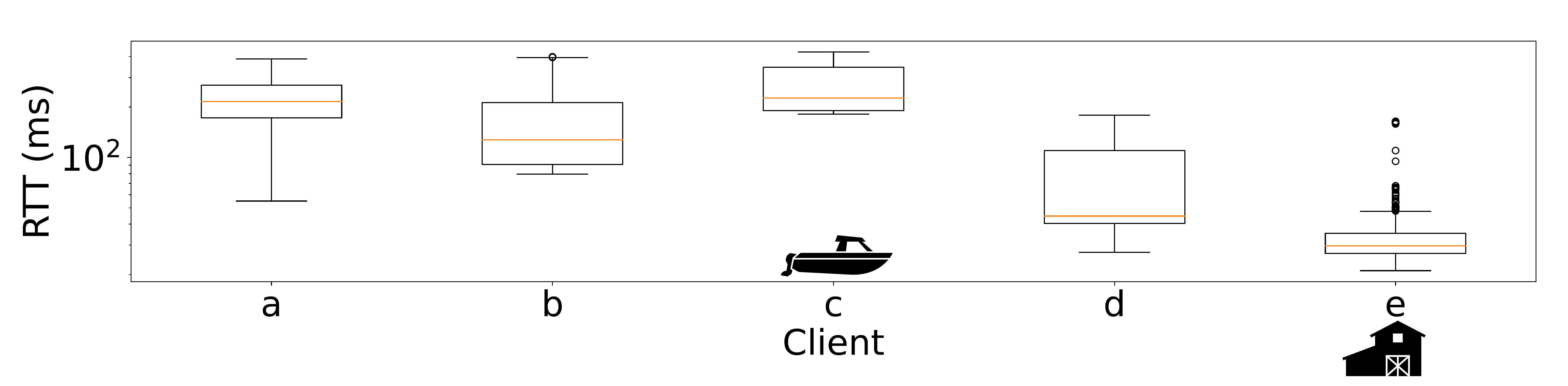}
    \caption{\textbf{Nigeria POP Per-Client Latency}}
    \label{fig:rtt_client_nigeria}
  \end{subfigure}
  %\hfill
    \begin{subfigure}{0.5\textwidth}
   \includegraphics[width=\linewidth]{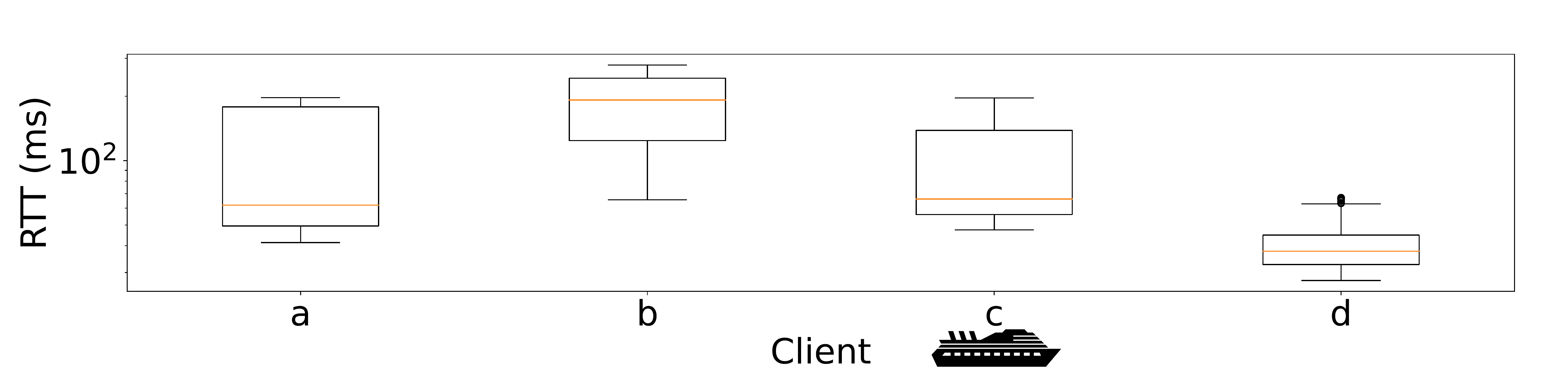}
  \caption{Brazil POP Per-Client Latency}
    \label{fig:wrtt_client_br}
  \end{subfigure}
    \begin{subfigure}{0.5\textwidth}
   \includegraphics[width=\linewidth]{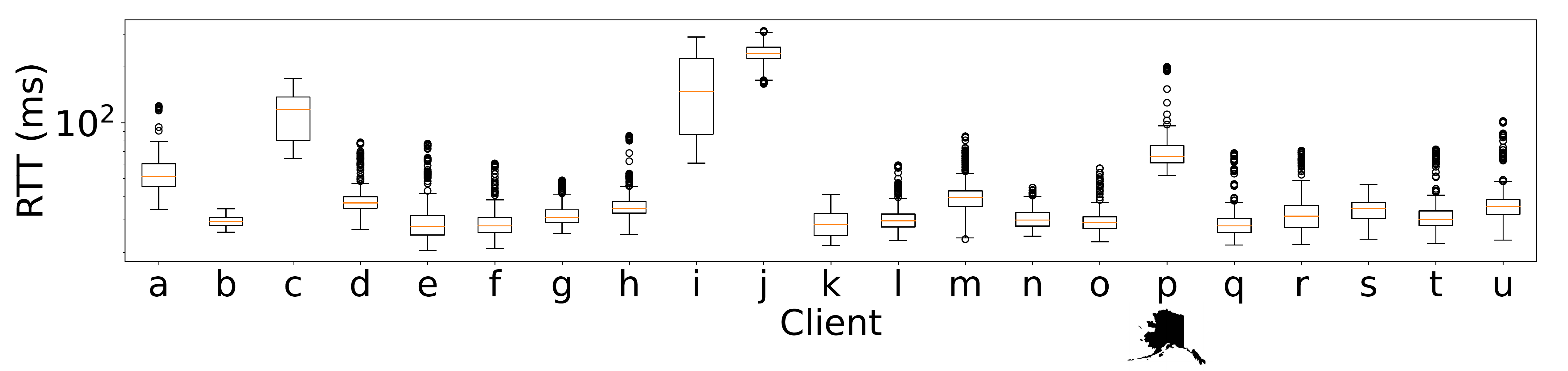}
  \caption{Washington POP Per-Client Latency}
    \label{fig:wrtt_client_wa}
  \end{subfigure}
      \caption{\textbf{Client Latencies (Log Scale)}---
      \textnormal{Due to a variety of factors (distance from POP, ISL usage), clients assigned to the same POP experience latencies that differ by an order of magnitude. }
      %\todo{make boats same} \todo{client->customer} }
    }
\end{wrapfigure}

\subsection{POP Distance}
\label{sub:sub:sec:pop_dis_min}

In this section, we investigate how a customer's distance to POP correlates with their expected latency.
We find that even with Starlink's vast network topology, remote customers can experience latency increases by over three fold compared to other customers assigned to the same POP.

% As shown in Section~\ref{sub:sec:hypatia_eval}, in simulation, a customer's minimum RTT exemplifies when a customer is connected to the ground station that is closest to the POP.  
% To test if the latency vs POP distance correlation holds in practice, we use 
In Figure~\ref{fig:latency_v_pop}, we plot the \lhh-measured latency and distances of the RIPE Atlas probes with their respective POPs. 
The further a Starlink customer is from their POP, the greater their minimum RTT.
In the worst case, a customer from the US Virgin Islands is assigned to their nearest POP, in Atlanta, Georgia, located roughly 1600~miles away.
They experience a minimum RTT nearly twice as large compared to customers that are closer to their own POP. 
% In Appendix~\ref{fig:latency_v_pop},  we show that average and maximnum RTT is not correlated with POP distance, and explain why this is the case in the next subsections (i.e., ISL usage). 

Minimum RTT can be used to approximate customer location when no ground truth is available (i.e., when measuring non-RIPE Atlas exposed services). 
For example, in Figure~\ref{fig:rtt_client_nigeria}, minimum RTT from data collected on May 18, 2023, indicates that customer \textit{e} must be located much closer to the Nigerian POP than customer \textit{c}. 
Indeed, customer \textit{c} is on a yacht near Seychelles (they host a TLS certificate registered to the name of a unique sportfisher yacht, which MarineTraffic.com shows to be near Seychelles~\cite{seychelles_ship})  while  customer \textit{e} is in the Nigerian Palm-Oil farm (i.e., the customer hosts a TLS certificate that fingerprints to a Nigerian Palm-Oil farm). 
Additionally, in Figure~\ref{fig:wrtt_client_wa}, we find that customer \textit{p}, whose minimum RTT is nearly 3~times larger than average minimum RTT, is thousands of miles away from their assigned nearest Seattle POP (customer \textit{p}'s exposed email server belongs to Vuntut Gwitchin First Nation, a cultural site, which is located thousands of miles away Seattle near the northern border of Canada).

Notably, no matter a customer's distance to their POP, all of our Starlink traceroutes show that packets are \textit{always} tunneled to/from the customers assigned POP before traversing the public Internet\footnote{Packets are always first routed through the customer's assigned Starlink POP once they enter the Starlink network. We conduct Section~\ref{sub:sec:lv_v_gt}'s experiment in reverse (contact the geographically distributed servers from our dish), and find that when sending packets from a Starlink dish, packets are always routed through the customers assigned POP before leaving the Starlink network.}, thereby causing unavoidable latency.
Tunneling customers through a home gateway across all connections is not unique to Starlink, but rather also common in mobile networks~\cite{mandalari2018experience}.

%\subsubsection{ISL Impact on Latency}
\subsection{Routing Changes}
\label{sub:sub:sec:isl_prevale}

\begin{figure*}[t]
  %\hfill
    \begin{subfigure}[t]{0.32\linewidth}
    \includegraphics[width=\linewidth]{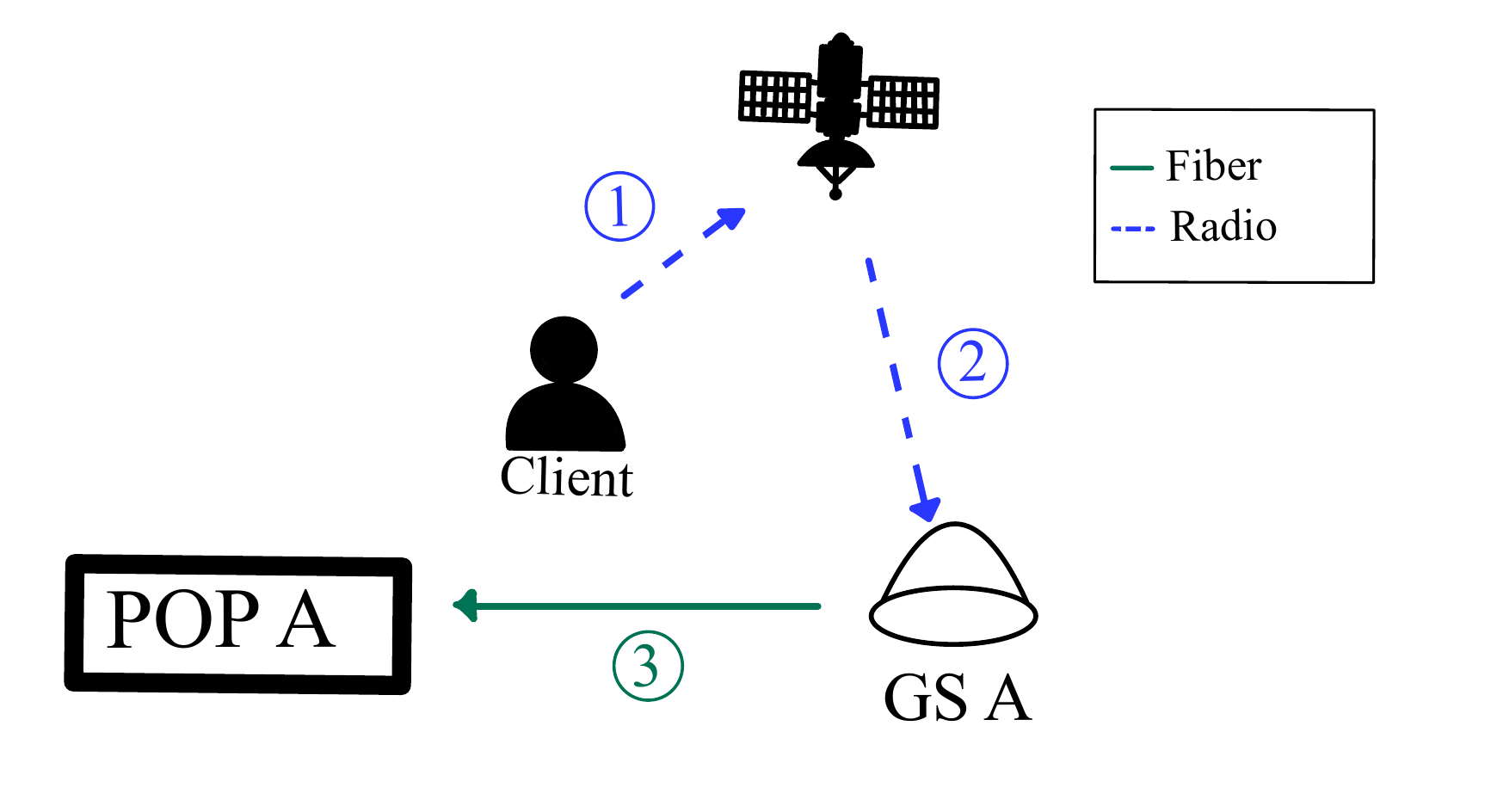}
    \caption{Single Hop Satellite Relay}
    \label{fig:client_no_isls}
  \end{subfigure}
  %\hfill
    \begin{subfigure}[t]{0.32\linewidth}
\includegraphics[width=\linewidth] {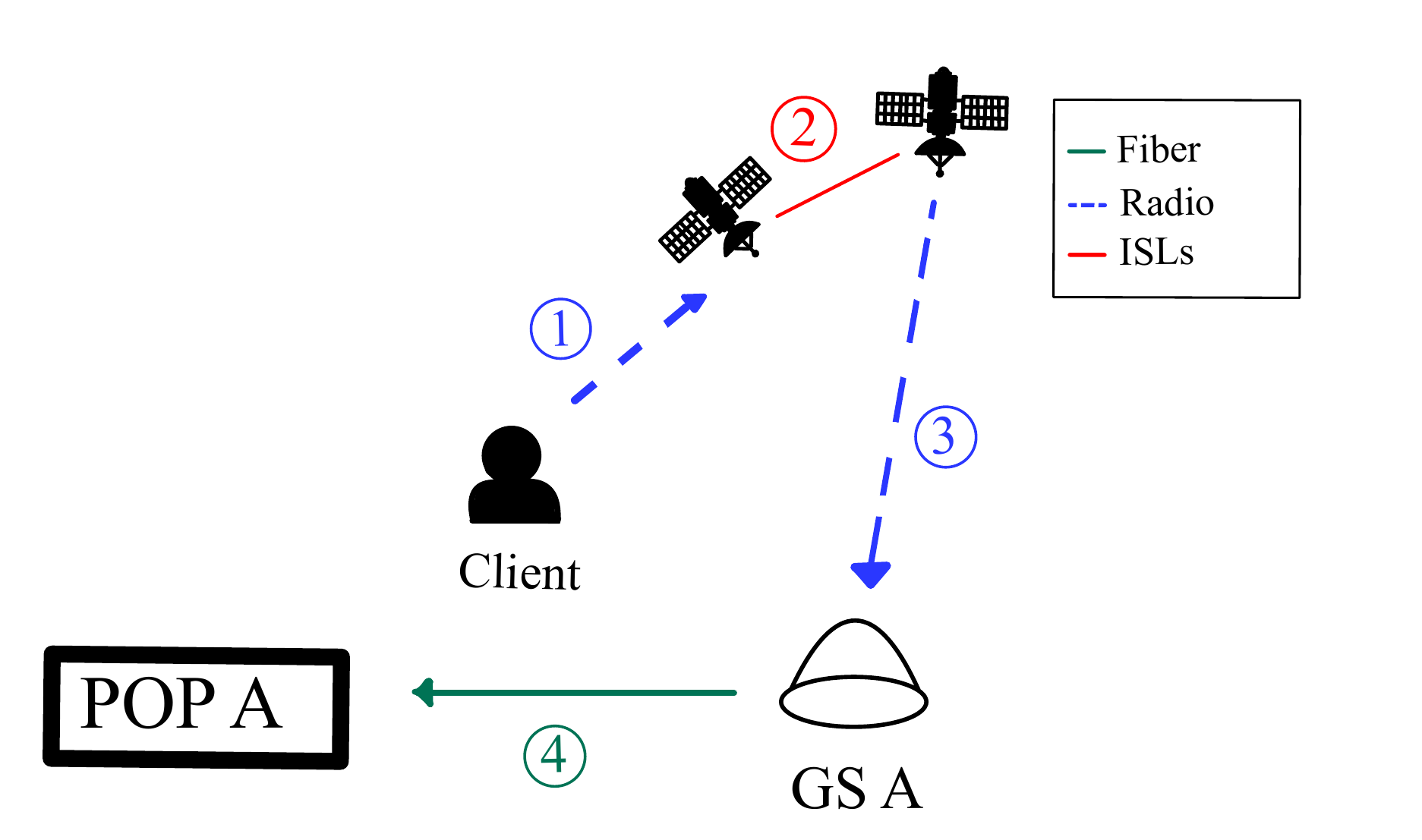}
  \caption{ISL Routing to Nearest \newline Ground Station}
    \label{fig:client_yes_isls}
  \end{subfigure}
  %\hfill
    \begin{subfigure}[t]{0.32\linewidth}
   \includegraphics[scale=0.25]{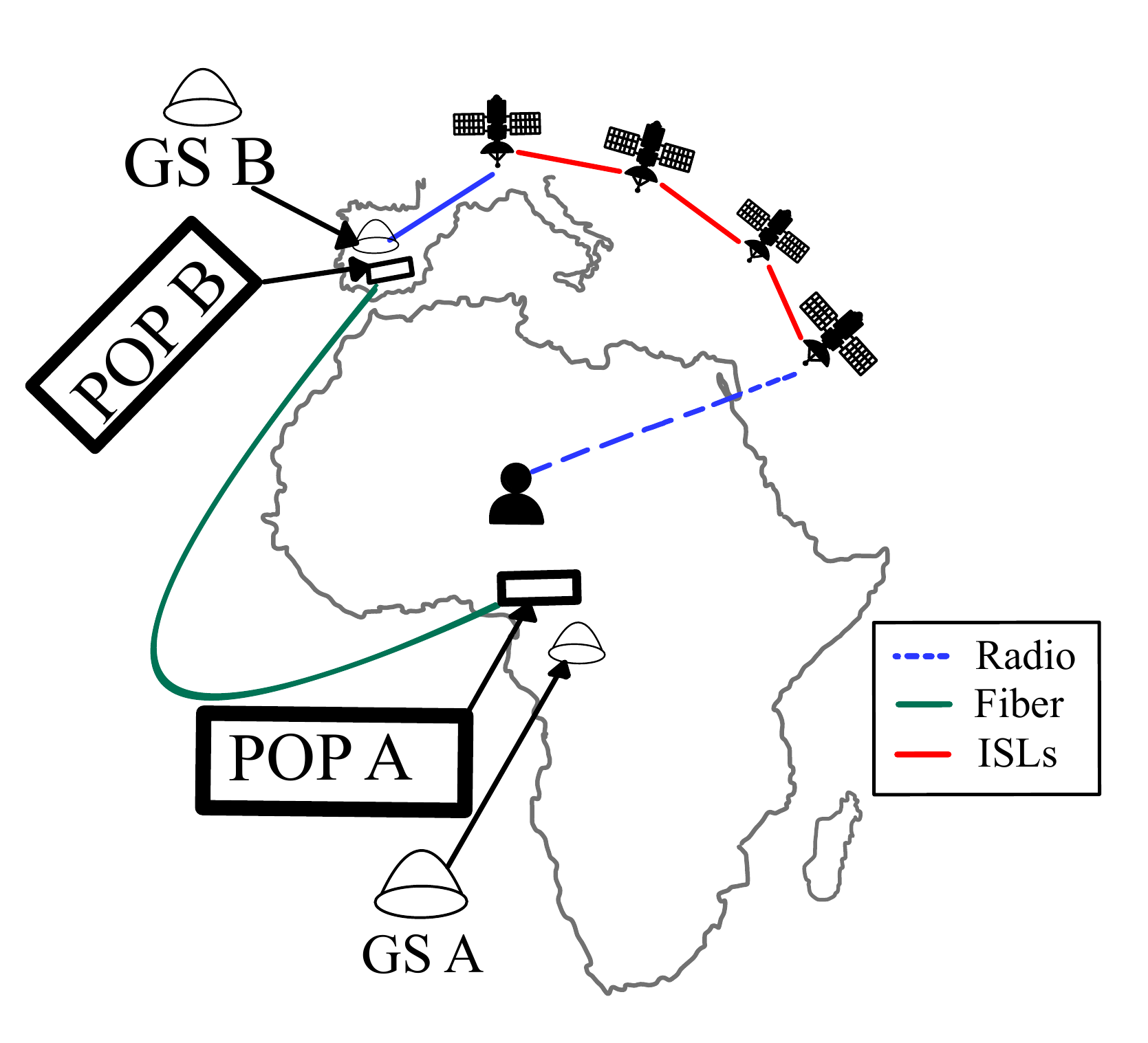}
  \caption{ISL Routing to Far \newline Ground Station}
    \label{fig:nigeria_client}
  \end{subfigure}
      \caption{\textbf{Routing Options}---
      \textnormal{Paths from a client to its POP can vary in satellite hops and length.}
    }
\end{figure*}

\begin{wrapfigure}{R}{0.5\textwidth}
  %\centering
    \includegraphics[width=\linewidth]{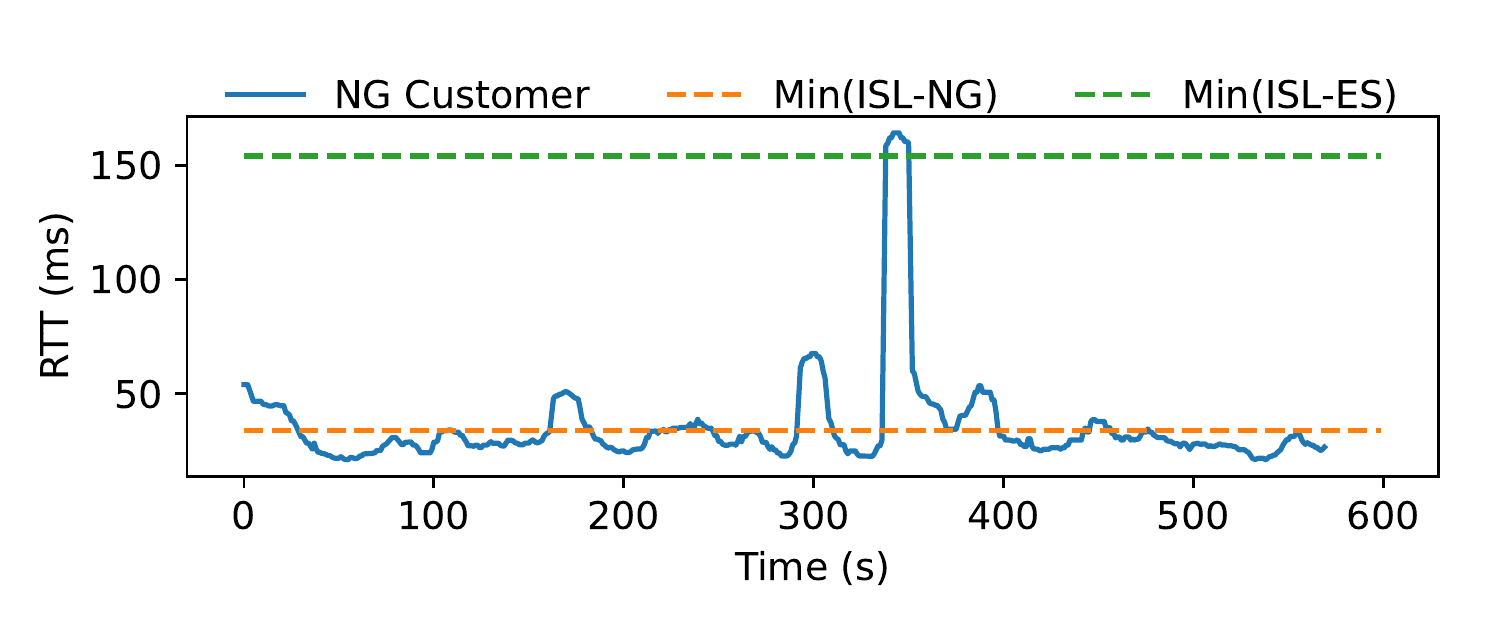}
    \caption{\textbf{ISL Impact On Latency Spikes}---% 
    \textnormal{Customer e located in Nigeria experiences spikes in latency that coincide with the expected latency of different ISL routing methods. We describe the different ISL routing methods in Section~\ref{sub:sub:sec:isl_prevale}. }}
    \label{fig:ng_isl_explain}
\end{wrapfigure}
% While POP distance and ISL reliance impact the best and average case RTT, we infer that outlier RTTs that last the entire satellite-connection period of 15~seconds (e.g., near second 350 in Figure~\ref{fig:ng_isl_explain}) is still due to the prevalence of ISL routing.
We next investigate how routing changes between a client, groundstation, and POP correlate with latency.
We find that customers physically located within ground station coverage experience surprising spikes in latency due to sub-optimal routing changes that are exacerbated by ISL technology, which Starlink engineers confirm. 

To deduce routing causes of latency, we take advantage of \lhh's ability to measure customer latency nearly anywhere in the world. 
We study a POP that minimizes the most variables in packet routing: the Nigerian POP.
The Nigerian POP has the least number\footnote{We rely on public crowd sourcing to collect country-specific SpaceX requests for ground stations~\cite{sat_earth_station_fcc,french,chile} to map the ground station topology of Starlink's network. Given Starlink's large fanbase who enjoy tracking ground station locations~\cite{starlinksx,unofficial_gs_map} we assume that our  knowledge is complete. Even if our knowledge was incomplete, sustained latency spikes that increase RTT by nearly an order of magnitude (e.g., second~350 in Figure~\ref{fig:ng_isl_explain}), could not be explained by connecting to nearby ground stations within one hop).} of ground 
stations (2) a negligible distance (80~miles) apart,
within a one-hop satellite distance.
We study a customer that we confirm is physically near their assigned Nigerian POP: the Palm-Oil customer (customer \textit{e} from Section~\ref{sub:sub:sec:pop_dis_min}).
Notably, without \lhh, one would have to solicit volunteers in Nigeria (of which currently there are none in RIPE Atlas) or travel with a dish to Nigeria. 

%https://www.reddit.com/r/Starlink/search/?q=find\%20ground\%20stations&restrict_sr=1
In the simplest case of routing, the Palm-Oil customer will connect to one of the two Nigerian ground stations nearby, using a one satellite ``relay'' hop (Figure~\ref{fig:client_no_isls}).
After connecting to any ground station, the packet is always \textit{terrestrially} routed to the POP, as revealed by Starlink~\cite{arcep,kentucky}\footnote{In Section~\ref{sec:continuous} we confirm that packets are routed terrestrially for as long as possible before reaching the satellite path. In Appendix~\ref{app:traceroute}, we confirm that no matter the public Internet destination, packets leave their assigned (nearest) pop and continue the route terrestrially.}.
To deduce how often relay routing occurs, we first plot the customer's RTT over time in Figure~\ref{fig:ng_isl_explain} from data collected on May 18, 2023. 
Second, we plot Hypatia's calculated minimum RTT when using non-relay routing (i.e., two satellites with ISLs) to the Nigerian ground station (``\texttt{Min(ISL-NG)}'').
Over 70\% of customer RTTs fall below \texttt{Min(ISL-NG)}, indicating that in the majority of cases the customer falls below the theoretical minimum latency of using two satellites to route to the nearest ground station, and therefore must be using single-hop relay routed to the nearest ground station.

However, one-third of the time, the Nigerian customer's RTTs increase between 2--5~fold the median, indicating that Starlink is likely connecting customers with distant ground stations\footnote{We record no packet drop during the RTT increases and know that a sustained latency increase is not ultimately due to a satellite switch (Section~\ref{sub:sec:obstruction_map}).}.
Indeed, Starlink engineers confirm that these sustained latency spikes are due to customers using ISLs, which route through ground stations located anywhere in the world---not necessarily in close geo-proximity\footnote{Starlink shares that routing is determined based on link availability, reliability, and capacity.}.
%The Nigerian customer's sustained latency spikes correlate with the expected latency of traversing longer routing paths.
%For example, 
Thus, the outlier latency (162~ms) near second~350 in Figure~\ref{fig:ng_isl_explain} could be the sum of the following footnoted equation\footnote{A Nigerian POP customer symmetrically routed through Spain would experience at least the following latency, where $c$ is the speed of light:
\begin{align}
\text{(Sat. RTT)}+\text{(ISL RTT NG to ES) } + \text{(Terr. RTT ES to NG)} \notag \\
%\end{align}
%\begin{align}
  = \text{(direct dist to sat)} + \text{(direct dist. NG to ES) } +  \text{(fiber dist. from ES to NG)}\notag \\
    = \text{(Hypatia calculated)} + \text{(distance / ISL speed) } +  \text{(ping test~\cite{ping-test})}\notag \\
%\end{align}
%\begin{align}
    = (11~\text{ms})*2 + (2200~\text{mi}/c)*2+  (110~\text{ms})\notag \\
%\end{align}
%\begin{align}
    = (11~\text{ms} *2)+(11~\text{ms}*2)+(110~\text{ms})=154~\text{ms.} \notag
\end{align}
}, which shows that it would take at least 154~ms for a Nigerian POP customer to route through the nearest non-Nigerian ground station, in Lepe, Spain (Figure~\ref{fig:nigeria_client}).
The Nigerian customer's terrestrial route from the Spanish ground station would span over 4970~miles, given that all customers traverse from their ground station to their assigned POP \textit{terrestrially} (Section~\ref{sub:sub:sec:pop_dis_min}).
% First/Last Hop Satellite RTT (calculated by Hypatia distance to reach first/last satellite to/from dish/groundstation), 
% ISL RTT to Spain (direct distance from Nigeria to Spain (8,000~KM) * speed of light) + terrestrial fiber RTT from Spain to Nigeria (submarine cable distance from Spain to Nigeria (9,000~Km)*speed of terrestrial fiber--2/3rds the speed of light~\tcite{} ) = 34ms+(18ms*2)+(45ms*2) = 160ms. 
The sustained latency spikes near seconds~175 and~300 in Figure~\ref{fig:ng_isl_explain} are likely due to ISLs routing back to the Nigerian ground station, as depicted in Figure~\ref{fig:client_yes_isls}. 
While the sustained peaks are above the theoretical minimum latency, when accounting for potential additional latency due to bad satellite selection\footnote{We compute worst case RTT using the methodology in Appendix~\ref{fig:worst_hypatia_star}.} (12ms) and an indirect ISL routing path\footnote{Starlink's first shell contains 22 satellites per orbit, creating a roughly 1243~mi distance between satellites. If using an extra satellite to ISL route through, that would create additional RTT of 2* 1243~mi/speed of light (13~ms).} (13~ms), the total RTT is within 5~ms of the second highest sustained peak at second~300.

While it may seem that always routing through a customer's assigned POP, even when a closer POP is available, is uniquely inefficient, routing through an assigned gateway is a wide-spread practice in IP Packet Exchange Networks (IPX) during international roaming.
Mandalari et al~\cite{mandalari2018experience} show that in IPXs, significant latency increases occur due to the strict requirement of packets routing through the customer's home gateway, no matter the packet destination, so that cellular carriers can easily monitor data usage, perform content filtering, etc. 
While Starlink \textit{customers} are often stationary, the satellites (and their routing path) are not, causing routing patterns similar to international roaming.
Coincidentally, the majority (70\%) of Starlink traceroutes leak MultiProtocol Label Switching, a protocol heavily used by IPXs to route between gateways~\cite{mandalari2018experience}.

We find that customers who must fully rely on ISLs experience latency that is substantially worse than the expected latency of direct ISL routing, which Starlink independently confirms~\cite{reddit_roam,reddit_moz}.
For example, the Seychelles-yacht customer thousands of miles away from Nigeria (Section~\ref{sub:sub:sec:pop_dis_min}) is surrounded by no ground stations, and therefore must solely rely on ISLs to route to its assigned Nigerian-POP. 
The Seychelles yacht's experiences a minimum RTT of 181~ms (6~times worse than Palm-Oil-Farm Customer's minimum RTT) indicating that it is theoretically improbable that ISLs are using a direct path to reach the ground station nearest to the Nigerian POP (i.e., a direct speed of light path from Seychelles to the Nigerian POP would take an RTT of 40~ms, and no more than 80~ms to account for potential ``zig-zag'' paths between satellites~\cite{hypatia}). 
Rather, the Seychelles yacht is also likely routed to a further ground station and suffers the additional latency of a terrestrial path back to the customer POP.
Making matters worse, the Seychelles yacht is likely frequently re-routed: during 42\% of our measurement period, the yacht's RTT is 150~ms over its minimum RTT.  

Sustained latency spikes are widespread and constant;
at least 70\% of customers experience at least one sustained latency every day during our month-long 5~minute data collection.  
Starlink engineers share that they are still building their ISL ``mesh'' and hope to improve edge case performance over time. 
In the next section, we find other instances of customers who experience impactful latency patterns and study how customer latency changes over time.

\subsection{Geographic and Temporal Patterns}
\label{sub:sec:geo_patterns}
\begin{figure*}[t]
  %\centering
    \includegraphics[width=\linewidth]{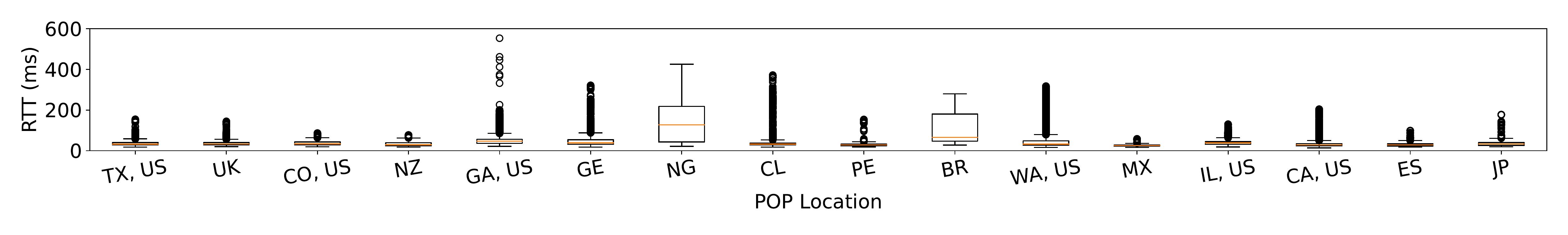}
    \caption{\textbf{Latency Across All Starlink POPs}---% 
    \textnormal{Customer latency varies dramatically depending upon their geographic location. Nigerian-POP customers experience the highest average RTT.}}
    \label{fig:rtt_box}
\end{figure*} 

\begin{figure*}[t]

    \begin{subfigure}{\linewidth}
  %\centering
    \includegraphics[width=\linewidth]{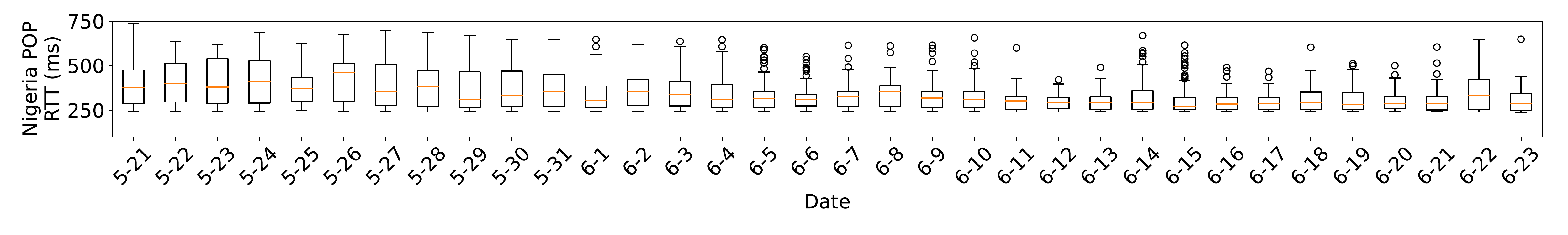}
    \caption{\textbf{Nigeria POP Latency Over Time}---% 
    \textnormal{Outliers in latency decrease over time.}}
    \label{fig:nig_overtime}
    \end{subfigure}
    \begin{subfigure}{\linewidth}
  %\centering
    \includegraphics[width=\linewidth]{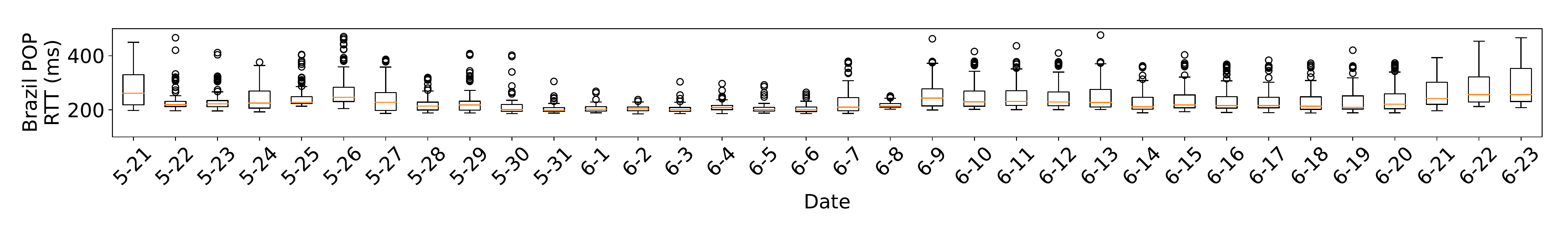}
    \caption{\textbf{Brazil POP Latency Over Time}---% 
    \textnormal{Outliers in latency briefly decrease before increasing again over time.}}
    \label{fig:br_overtime}
    \end{subfigure}
    \begin{subfigure}{\linewidth}
  %\centering
    \includegraphics[width=\linewidth]{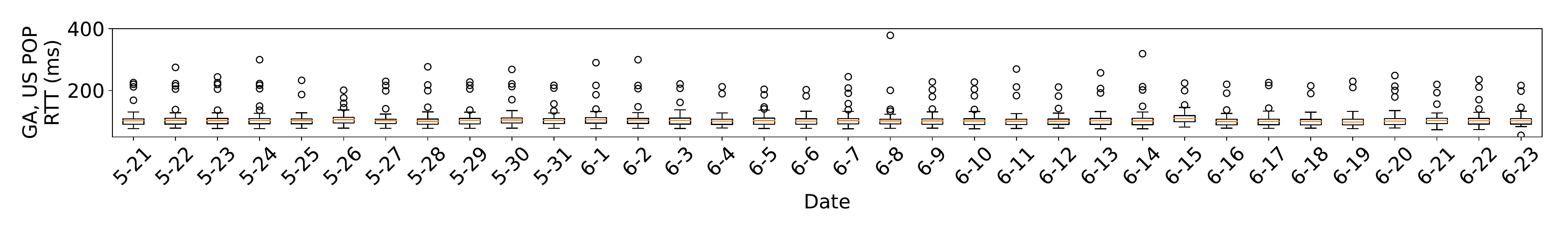}
    \caption{\textbf{Georgia, US POP Latency Over Time}---% 
    \textnormal{Latency remains stable over time.}}
    \label{fig:ga_overtime}
    \end{subfigure}
    
    \caption{\textbf{Latency Over Time}---POP latency changes over time depending upon the geographic location.}
\end{figure*} 

Latencies across POPs significantly vary due to their assigned types of customers (e.g., maritime, stationary) and distances of customers (e.g., extremely remote).
We plot the distribution of customer RTTs across all customers and POP locations in Figure~\ref{fig:rtt_box}.
Average RTTs vary by over 500\%: from 28~ms (Mexico)--149~ms (Nigeria). 
Standard deviation of RTTs also vary by an order of magnitude: from 4~ms (Mexico)--109~ms (Nigeria).
While customers with different RTTs are commonplace across all POPs (Figure~\ref{fig:rtt_client_nigeria}--Figure~\ref{fig:wrtt_client_wa}), we describe how certain geographies are more or less likely to attract unique customer patterns:

% Customers assigned to the same Starlink Point Of Presence (POP) experience different distributions of packet round trip times (i.e., latency).
% In Figure~\ref{fig:rtt_client_nigeria}, we show the distribution of RTTs for every customer with a reachable service assigned to the Nigerian POP. 
% While client `e' experiences an average RTT of 28~ms and a minimum RTT of 20~ms, client `c' experiences an average RTT of 250~ms and a minimum RTT of 200~ms.
% Different RTT distributions are not correlated with extremely different packet loss rates: client e loses 6\% of packets  while client c loses only 4\% (although it experiences higher RTTs). 
% Customers with different RTTs are commonplace across all POPs, including Brazil (Figure~\ref{fig:wrtt_client_br}) and Seattle (Figure~\ref{fig:wrtt_client_wa}).  
% In the next sections, we describe the causes behind different customer latency patterns. 

\vspace{3pt}
\noindent
\textbf{Nigeria.}\quad
As of October 2023, Nigeria contains the only POP for the entire continent of Africa, which causes African marine traffic (e.g., Seychelles yacht in Figure~\ref{fig:rtt_client_nigeria})
to be routed through the Nigeria POP. 
Since marine traffic must rely on ISLs, which correlate with high average RTTs (Section~\ref{sub:sub:sec:isl_prevale}), it is no surprise that Nigeria customers experience the worst RTTs. 
Nevertheless, we see Nigeria customer latency decrease overtime. 
In Figure~\ref{fig:nig_overtime}, we show that between May--June, 2023 the population of customers that experience outlier behavior shrinks and median latency decreases by 25\%, from roughly 400~ms to 300~ms. 
The downward trend in latency provides encouragement that Starlink is actively changing routing patterns.

\vspace{3pt}
\noindent
\textbf{Brazil.}\quad
Brazil-POP customers experience the second largest average RTT (104~ms) and standard deviation of RTT (27~ms).
While Brazil has at least~10 ground stations, its coastal location also attracts marine traffic that relies on ISLs, which correlate with substantially increased latency.
For example, customer Oceanica Sub VII (identified using an exposed SNMP firewall name) who is off the coast of Fortaleza Brazil~\cite{vesselfinder} at the time of our experiment is 1491~mi away from its POP---a 14~ms RTT if using a direct ISL path---but experiences an average RTT of 96~ms (Figure~\ref{fig:wrtt_client_br}), indicating indirect routing is in use. 
Consistent with other marine traffic, Oceanica Sub VII continually experiences sustained latency spikes, lasting for 16\% of our measurement period. 
Brazil customers do not see overall latency improvement in the same way as Nigerian customers do (Figure~\ref{fig:br_overtime}).

%SNMP firewall name, ``OCEANICASUBVII,'' fingerprints the customer aboard the Oceanica Sub VII, which is a  vessel off the coast of Fortaleza Brazil~\cite{vesselfinder} at the time of our experiment. 

\vspace{3pt}
\noindent
\textbf{Georgia, USA.}\quad
Georgia-POP customers experience the highest average (58~ms) and standard deviation (21~ms) of latency across the US. 
Ground truth location from RIPE atlas probes indicate that on average, a Georgia-POP customer is 1,000~miles away, partly due to customers located in the US Virgin Islands being assigned the Georgia POP. 
Thus, customers' often long distance to the Georgia-POP increases the lower bound of the latency that the majority of customers experience. 
Georgia latency patterns remain relatively stable over time (Figure~\ref{fig:ga_overtime}).

\vspace{3pt}
\noindent
\textbf{Other POPs.}\quad Peru, Australia, and New Zealend experience the shortest RTT times, on average under 30~ms. 
All three locations surrounded by at least six groundstations that are reachable by a single satellite-hop. 
We do not identify any maritime customers assigned to those POPs.

\subsection{Summary}
\lhh's global perspective shows how Starlink's complex network architecture impacts customer latency across many facets. 
First, we find that depending upon a customer's distance to their POP, minimum latency grows over three-fold.
Second, we detect and validate that routing changes to different ground stations cause sustained latency spikes that increase RTT by near an order of magnitude. 
Third, reliance on ISLs (e.g., customers on boats) correlates with increases in RTT and sustained latency increases.
Notably, the diversity of customers and routing infrastructure causes large variances in RTT across POPs worldwide,
underlining the value of the global and diverse perspective provided by \lhh.

\section{Limitations and Future Work}
\label{sec:future}

\lhh does not replace existing methodologies, but rather serves as a low-barrier methodology that provides data about LEO links worldwide. 
We see many possible directions to integrate \lhh into future LEO research:

\vspace{3pt}
\noindent
\textbf{Measuring Other LEO Networks.}\quad
While Starlink is the only LEO satellite network that sells to individuals, other LEO networks  (e.g., AWS Kuiper, Telesat, etc) are expected to provide new services in the coming years. 
With different networking architectures, including differently-arranged satellite constellations, these networks will likely exhibit both similar and different behaviors to Starlink. 
\lhh can be applied to measure other LEO networks once customers begin to expose services.
\lhh collected data can illuminate the similarities and differences of LEO network performance across different architectures in practice. 

As an example, we use \lhh to identify roughly 20~OneWeb measurable-endpoints that belong to business customers located in primarily Northern regions (e.g., Alaska, Canada).
We detail our exact OneWeb-specific \lhh methodology in Appendix~\ref{sec:oneweb}.
Unfortunately, we are not able to validate latency measurements with ground truth (i.e., a OneWeb dish we control) because OneWeb dishes are (1) only available to businesses and (2) prohibitively expensive (e.g., upwards of \$23,000~\cite{oneweb_cost}).
We hope future work with access to OneWeb equipment can validate our OneWeb-\lhh methodology.

\vspace{3pt}
\noindent
\textbf{Measuring LEO Networks Over Time.}\quad
As Starlink continues to add more customers, ground stations, POPs, satellite orbits and routing policies, \lhh can be used to compare how the architecture changes network measurements over time. 
For example, how does adding more users in a single location affect congestion, bandwidth and latency?
Do different routing policies fundamentally change customer network quality of experience?
%How does the adoption of ISLs fundamentally change customer network quality of experience?
We are open sourcing \lhh daily-collected data, giving researchers immediate access to answer temporal questions. 

\vspace{3pt}
\noindent
\textbf{Data-Driven Simulations.}\quad
While simulations allow for a wider flexibility of experiments than measurements, simulation accuracy is constrained by a plethora of unknowns about how real LEO networks operate. 
%For example simulations can test new congestion control algorithms, routing schedules, application layer behaviors, etc. 
Future work should look into 
(1) training predictive models with \lhh collected data to better simulate real-world networking conditions world wide,
(2) creating ``replay'' models with \lhh data that can test the performance of new algorithms (e.g., congestion control) using real data from the past.

% \vspace{3pt}
% \noindent
% \textbf{Geo-location for LEO Exposed Services.}\quad
% While Starlink's IP Geo-location feed~\cite{starlink-loc} provides city granularity, its granularity is not tailored to track roaming users (Section~\ref{sub:sub:sec:pop_dis_min}). 
% Future work should develop automated methods to identify granular locations for exposed services. 
%For example, since RIPE atlas provides ground truth locations for probes, one can use those to develop a \lhh LEO-specific methodology to estimate geographic location. 

\vspace{3pt}
\noindent
\textbf{LEO Network Coverage.}
While \lhh provides over 10~times more coverage of a LEO network than the leading alternative~\cite{ripe_atlas}, \lhh does not provide full coverage.  
For example, while Starlink has an estimated 2~million users~\cite{spaceX-2mil}, \lhh only measures an estimated 0.1\% of all customers.
Furthermore, \lhh is biased towards measuring customers who host exposed services, which may introduce confounding factors. 
Future work should investigate if other opportunities exist to measure LEO customers that do not expose services, to further increase coverage and reduce bias. 
\section{Conclusion}

In this work we introduced \lhh, a methodology for measuring LEO satellite networks at scale. 
\lhh builds on the observation that Internet exposed services that use LEO-based Internet access can reveal both satellite network architecture and performance, without needing physical hardware.
\lhh is accurate and provides an order of magnitude more coverage than alternative solutions.
Using our new global perspective, we study over 2.4K Starlink customers across farms, boats, and remote regions, to understand user latency. 

Our investigation surfaces that contrary to prior assumptions, sustained peaks of latency are not caused by distant satellite location.
We highlight that ISL routing patterns create the widest variance of latency.
While ISLs were advertised as a low latency solution for more direct routing~\cite{hypatia,musk_isls},
they significantly increase the length of the routing path between the ground station to POP, at the benefit of increasing connectivity (e.g., to ships).

Connectivity at the cost of latency is not unique to Starlink; mobile networks face the same trade-off when providing international roaming under different regulatory bodies~\cite{mandalari2018experience}. 
By increasing connectivity through ISL deployment, Starlink customers now occasionally experience RTTs that are inching closer to GEO latency.
%A customer's average latency today ranges from 20~ms to over 150~ms
% High RTTs stand in the way of LEO's primary selling point of low latency, especially since GEO-stationary satellite networks have always provided vast coverage and high bandwidth~\cite{viasat-3}.
As LEO networks continue to increase connectivity, we hope the community uses \lhh to understand their real-world deployment when continuing to help design and protect the LEO ecosystem. 

\newpage
{
\balance
\bibliographystyle{ACM-Reference-Format}
\bibliography{reference}
\appendix
%\onecolumn
\pagebreak

\section{Ethics}

Our work does not involve human subjects and therefore, according to our institution's IRB policies, does not require IRB approval. 
Nevertheless, we agree with and support the mission of minimizing harm when measuring LEO links, and thoroughly discuss the measures we take in  Section~\ref{sub:sec:ethics}. 

% \section{How LEO Services Become Exposed}
% \label{app:how_expose}

% To configure a service to be publicly accessible over the Internet, the service requires a publicly routeable IP address. 
% We find that different customers must jump through different hoops to expose services over LEO satellite networks.  

% For example, while all Starlink customers on the ``Business'' plan\footnote{\url{https://www.starlink.com/business}} can request both a publicly routeable IPv4 and IPv6 address,  customers on all other plans can only request publicly routeable IPv6 addresses. 
% Nevertheless, there have been cases where non-business Starlink users unsolicitedly receive public IPv4 addresses from Starlink~\cite{reddit_starlink_1,reddit_starlink_2,reddit_starlink_3}.  
% The current generation of Starlink routers do not allow port-forwarding, requiring users to use their own routers to host publicly accessible services~\cite{reddit_ddwrt}. 

% Since OneWeb only caters to enterprise clients and governments, their exists little public information regarding the specifics of their networking solutions. 
% Nevertheless, their dedication to custom solutions~\tcite{}, likely includes providing public IP addresses and port-forwarding ability to clients. 

\section{Mapping The Starlink Network With Traceroute }
\label{app:traceroute}
To map Starlink, we first experiment with different tools and protocols, to determine which illicit the most network information. 
We experiment with running ICMP, UDP, and TCP traceroutes across the following tools, which differ in the manner they construct and send packets: dublin-traceroute~\cite{dublin_tr},  paris-traceroute~\cite{paris_tr}, tcp-traceroute~\cite{tcp_tr}, mtr (with the mpls flag)~\cite{mtr_tr}, and TNT~\cite{vanaubel2019tnt}. 
When scanning all Starlink customer endpoints (Section~\ref{sec:continuous}) from \Stanford, 60.3\% are reachable across all tools using ICMP, 17.2\% are reachable using UDP, and 24.6\% using TCP. 
The number of routing hops within the Starlink network nearly always (96\% of the time) does not change depending upon the tool or protocol used.

After performing all variations of traceroute, we find three noteworthy network characteristics of Starlink:
(1) Starlink routes TCP and UDP packets through a changing set of IP addresses before reaching the endpoint, while keeping a consistent routing path for ICMP packets.
While 98.3\% of tcptraceroutes result in a second-to-last hop set of at least two IP addresses, 100\% of ICMP paris-traceroutes result in in a second-to-last hop set of just one IP address. 
(2) Starlink's consistent routing path traverses the customer's POP across 100\% of traceroutes, no matter from/to where the server and client are. 
%As a result, tools that do not have TCP capabilities, such as dublin-traceroute and paris-traceroute are less visible to some internal networking paths of Starlink.
(3) Starlink uses MPLS routing before reaching the end-host. 
Consequently, a subset of routing is not visible to any traceroute tool. 
TNT, a tool used to uncover routing behind MPLS does not find any new networking paths within Starlink. 
While mtr reveals the last label assigned to a packet at the end of an MPLS tunnel, we do not identify any useful patterns.

\subsection{Mapping Starlink's Routing}
\label{app:sub:map_internal}
\label{sub:sec:isolate_sats}

To obtain ground truth on Starlink's internal network operation, we purchase a Starlink generation 2 router and dish~\cite{starlink_residential}, and deploy it in \SD.
To identify where internal satellite routing occurs, we conduct egress ICMP paris-traceroutes from our dish to a diverse set of end-points, including at least 3 LEO satellite endpoints (Section~\ref{sub:sec:coverage}) in every available country, all the geographically-distributed DNS root servers, and geographically distributed AWS servers (described in Section~\ref{sec:eval}).
%In Figure~\ref{fig:alg_1}, we show an example egress traceroute from our dish to the \Stanford-server, and in Figure~\ref{fig:alg_2} we show an ingress traceroute from the \Stanford-server to our dish. 
Across all egress traceroutes, once a packet leaves our dish's local area network (LAN), there is a spike in the round trip time (e.g., 39ms in Table~\ref{fig:alg_1}).
All subsequent Starlink (ASN\,14593) hops incur negligibly different round-trip-times, suggesting that the first hop encompasses at least the entire satellite link (i.e., dish to satellite to groundstation). 
We further validate in Section~\ref{sec:eval} that the satellite link is within the first hop. 

% between the dish and the first non-dish-LAN hop that is always assigned the IPv4/IPv6 address 100.0.64.1/2605:59c8:3000:f27f::1
% \footnote{Traceroutes conducted by RIPE Atlas Starlink Nodes and traceroutes posted on Reddit also confirm that the first non-client-LAN hop traverses 100.0.64.1/2605:59c8:3000:f27f::1}. 

\begin{wraptable}{R}{0.5\textwidth}
%\centering
\small
\centering
\begin{tabular}{llll}
\toprule
Hop & Router IP & RTT & Network \\
& & (ms) & \\
\midrule
 1 & 2605:59c8:3049:fa00::1 & 1  & Dish (LAN) \\ %192.168.1.1
 2 & 2605:59c8:3000:f27f::1&39   & Starlink \\ %100.64.0.1
 3 & 2620:134:b0fe:251::114& 38   & Starlink \\ %172.16.251.114
 4 & 2620:134:b0ff::378 &  38  & Starlink \\ %206.224.65.254 
 5 & 2620:134:b0ff::368 & 38    & Starlink\\ %206.224.65.234
 6 & 2620:107:4008:d03::1 &38  & Cogent \\ %198.32.251.6
 %7 & Continued... & ... & ... \\
\bottomrule
\end{tabular}
\vspace{8pt}
\caption{\textbf{Truncated Traceroute From Dish to Public Server}---%
\textnormal{ 
The LEO link is traversed between the first and second hop, as indicated by the single spike in latency. }}
\label{fig:alg_1}
%\vspace{-15pt}
\end{wraptable}

While the first visible non-LAN hop includes the satellite link, it likely also includes the terrestrial pathway between the groundstation and POP. 
% Starlink's network configuration allows a dish owner to measure the path between a dish and a POP, but it does not allow the detection of ground station. 
Starlink addresses that are one hop away from a different autonomous system (e.g., hop 5 in Table~\ref{fig:alg_1}) likely belong to equipment located in Internet Exchange Points (i.e., Starlink PoPs), as they are  
(1) often  (68.2\% of the resolvable hostnames) preceded by an IP address a hostname suffix that indicates an IXP presence (e.g., any2ix.coresite.com, ch3.unitedix.net)
(2) nearly always (87.6\%) experience less than 1ms RTT difference between the preceded router. 
All Starlink IP addresses between the POP hop and the client incur negligible additional latency, indicating that they are likely not a ground station, which, in our experiment, is built at least 100~miles away from the POP. 
%Debug statistics provided by client dishy hardware also report latencies to the dish's POP, but not to the groundstation~\tcite{}.
Thus, with physical equipment and today's networking tools, Starlink's satellite link will include terrestrial latencies between the groundstation and POP. 
We further evaluate the groundstation impact in Section~\ref{sub:sec:groundstations}.

% \todo{OneWeb Window size = 10}

\section{\lhh Implementation Details Continued}
\label{app:manda_ping_details}
%\todo{ }
%Integrate Manda's notes here
%https://docs.google.com/document/d/1YcrnDb0Mcy3OhxU83gk4NQVnwVSpiFz3Oa2Klp9Xo0o/edit#

We define a ttl ping as an ICMP paris-traceroute with the first hop and max-ttl set to the same hop number. To ping the terrestrial router we set both the first hop and the max-ttl to the terrestrial-hop-router-hop number (i.e., the second-to-last visible hop in the paris-traceroute). To ping the exposed service router we set both the first hop and the max-ttl to the exposed-service-hop-router-hop number (i.e., the last hop in the paris-traceroute). We opt to use ttl pings over regular pings due to the lack of coverage regular pings provide for terrestrial routers. We find that 99.2\% of regular pings to terrestrial-hop-router-hops never respond. In contrast, 0\% of ttl pings to terrestrial routers resulted in complete packet loss.

To use ttl pings, we rely on the assumption that the terrestrial-hop-router-hop number and the exposed-service-hop-router-hop number and the IP addresses they map to are stable. We validate the assumption by running 100 paris-traceroutes, one second apart, to each exposed service. We find that the vast majority of the exposed-service-hop-router-hop numbers (96.1\%) and terrestrial-hop-router-hop numbers (99.9\%) are consistent across the ICMP paris-traceroutes to the same exposed service. Additionally, we find that IP addresses for the same hop numbers do not change across ICMP paris-traceroutes.

\section{Using a Starlink Dish Obstruction Map To Infer Satellite Location}
\label{app:obstruction_meth}
Starlink's satellite dish does not directly reveal which satellite it is connected to.
However, we notice that Starlink's obstruction map---a map that highlights where the dish's visibility is obstructed---records the location of the satellite that is connected to the dish. 
To infer which satellites the Starlink dish is connected to during our experiment, we first reboot our dish to clear the dish's cached obstruction map and visually validate that the dish's obstruction map is clear. 
We then start our experiment and send the ``get\_obstruction\_map'' gRPC command every second to receive the obstruction map every second. 
After reboot, we notice a 30~second delay before ``get\_obstruction\_map'' returns data.
Once our experiment is finished, we subtract the obstruction map at time $t-1$ from the obstruction map at time $t$, to illuminate the location of the dish-connected satellite is at time $t$.  
Visual validation shows that every second the dish connects to a new satellite location that is either near the original one (i.e., the same satellite is connected to the dish) or at a completely different location (i.e., a new satellite is now connected to the dish).

\section{Hypatia Worst Case Satellite Selection Predictions}
\label{app:eval}

We modify Hypatia's source code such that, when simulating routing patterns, it chooses the worst-case option for satellite routing (i.e., it connects the client with satellite that maximizes its round trip time to the ground station.
We additionally configure Hypatia to reflect the azimuth properties of a Starlink dish. 
Since a Starlink dish cannot connect to satellites behind itself~\cite{starlink_FOV}, and points itself at an azimuth of -22 N, we configure Hypatia to choose satellites for the dish that are between an azimuth lower than $180-22=158$ or greater than $360-22=338$. 
The ground station is not configured with an azimuth cut-off, as their antennas can connect to satellites across all 360~degrees. 
In Figure~\ref{fig:worst_hypatia_star}, we illustrate the worst case predictions between the client and a subset of ground stations.
Unfortunately, even worst case predictions do not capture Starlink's RTT dynamics.

\begin{wrapfigure}{R}{0.5\textwidth}
  %\hfill
      \centering
   \includegraphics[width=\linewidth]{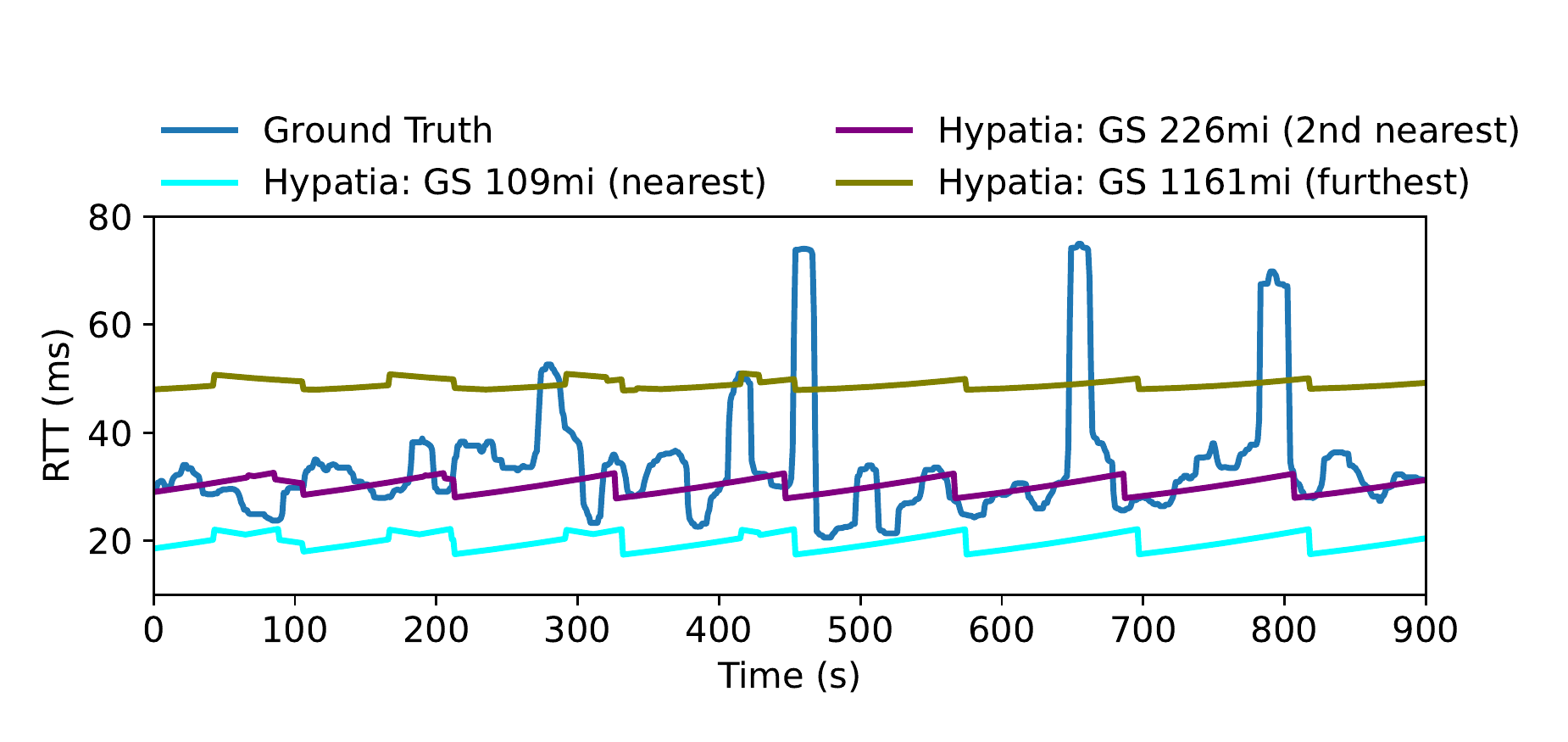}
  %   \begin{subfigure}{\columnwidth}
  %       \centering
  %   \includegraphics[width=0.5\columnwidth]{figures/pot_hypatia_anon_all.pdf}
  %   \caption{\textbf{Hypatia Best--Worst Case Predictions Across All Ground Stations}}
  %   \label{fig:hypatia_anon}
  % \end{subfigure}
  %\hfill
      \caption{\textbf{Worst Case Hypatia Predictions}---
      \textnormal{Simulations do not capture the dynamics of real-world Starlink RTTs.}
    }
    \label{fig:worst_hypatia_star}
\end{wrapfigure}

\section{Using \lhh to Measure OneWeb}
\label{sec:oneweb}

OneWeb is a LEO constellation that provides Internet for businesses, rather than individual customers~\cite{oneweb}. 
We apply the \lhh methodology to measure OneWeb.
However, we are not able to validate latency measurements with ground truth (i.e., a OneWeb dish we control) because OneWeb dishes are (1) only available to businesses and (2) prohibitively expensive (e.g., upwards of \$23,000~\cite{oneweb_cost}).

We outline the OneWeb-specific \lhh steps below.

\vspace{3pt}
\noindent
\textbf{1. Collect Exposed Services.}
We use Censys to collect all exposed services in the 
OneWeb network (AS\, 800).

\vspace{3pt}
\noindent
\textbf{2. Filter for Customer Endpoints.}
For OneWeb services, customer IP addresses resolve to business names. 
Thus, \lhh uses a blocklist approach in which it filters for PTR or SOA records that do not contain OneWeb's domain, nor the domain of any regional Internet registry (i.e., AFRINIC, ARIN, APNIC, LACNIC, or RIPE).

\vspace{3pt}
\noindent
\textbf{3. Exclude PEPs.}
We do not identify any PEPs.

\vspace{3pt}
\noindent
\textbf{4. Geolocate services.}\quad
To determine the geographic location of OneWeb services, we leverage an observation: OneWeb enterprise customers are listed as the registrant of One Web IP addresses in the whois database.
Thus, we use the location of the registrant to determine the likely location of the service. 
Unlike Starlink, the majority of OneWeb expoesed services are located in Northern regions, including Alaska. 
%\todo{add one web to table}

\vspace{3pt}
\noindent
\textbf{4.Identify satellite-routed path.}\quad
OneWeb does not publicly reveal internal network operations.
Moreover, since OneWeb only caters to enterprise and government clients, we do not have access equipment that can provide an internal network perspective. 
We conduct Ingress traceroutes to all the OneWeb endpoints from a server in \Stanford, and show an example traceroute in Table~\ref{fig:alg_oneweb}.
To measure OneWeb satellite latency, we subtract the second-to-last visible hop from the last hop.

\begin{wraptable}{R}{0.5\textwidth}
%\centering
\small
\centering
\begin{tabular}{llll}
\toprule
Hop & Router IP & RTT & Network \\
& & (ms) & \\
\midrule
 3 & 104.255.10.149& 50   & Astute Hosting \\ 
 4 & *&    &  \\ %206.224.65.254 
 5 & * &    &  \\
 6 & Customer IP &118  & OneWeb \\ %198.32.251.6
\bottomrule
\end{tabular}
\vspace{8pt}
\caption{\textbf{Truncated Traceroute From Public Server to OneWeb}---%
\textnormal{ 
The LEO link is traversed between the last and second-to-last hop, as indicated by the single spike in latency. }}
\label{fig:alg_oneweb}
%\vspace{-15pt}
\end{wraptable}

}

%ACK
%Nitinder
%SpaceX people: Nathan + Maxim

\end{document}